\documentclass[journal]{IEEEtran}
\usepackage{amssymb,amsmath,amsfonts,bm, graphicx,theorem,latexsym}
\usepackage{rotating,setspace,latexsym,color,subfigure}
\usepackage{cite,authblk}
\usepackage{url}

\begin{document}
%\IEEEoverridecommandlockouts

\title{Energy Harvesting Wireless Communications: \\ A Review of Recent Advances}

\author{S. Ulukus, A. Yener, E. Erkip, O. Simeone, M. Zorzi, P. Grover, and K. Huang

\thanks{Manuscript received April 1, 2014; revised September 19, 2014; accepted November 10, 2014. The associate editor coordinating the review of this paper was L. Milstein.}

\thanks{S. Ulukus is with the Department of Electrical and Computer Engineering, University of Maryland, College Park, MD 20742 USA (e-mail: ulukus@umd.edu).}

\thanks{A. Yener is with the Department of Electrical Engineering, Pennsylvania State University, University Park, PA 16802 USA (e-mail: yener@ee.psu.edu).}

\thanks{E. Erkip is with the Department of Electrical Engineering, Polytechnic Institute of New York University, Brooklyn, NY (email: elza@poly.edu).}

\thanks{O. Simeone is with the Department of Electrical and Computer Engineering, New Jersey Institute of Technology, Newark, NJ 07102 USA (e-mail: osvaldo.simeone@njit.edu).}

\thanks{M. Zorzi is with the Department of Information Engineering, University of Padova, Via G. Gradenigo 6/B, 35131 Padova, Italy (email: zorzi@dei.unipd.it).}

\thanks{P. Grover is with the Department of Electrical and Computer Engineering, Carnegie Mellon University, Pittsburgh, PA 15213 USA (email: pgrover@cmu.edu).}

\thanks{K. Huang is with the Department of Electrical and Electronic Engineering, University of Hong Kong, Hong Kong, China (e-mail: huangkb@ieee.org).}
}

\maketitle

\begin{abstract}
This article summarizes recent contributions in the broad area of energy harvesting wireless communications. In particular, we provide the current state of the art for wireless networks composed of energy harvesting nodes, starting from the information-theoretic performance limits to transmission scheduling policies and resource allocation, medium access and networking issues. The emerging related area of energy transfer for self-sustaining energy harvesting wireless networks is considered in detail covering both energy cooperation aspects and simultaneous energy and information transfer. Various potential models with energy harvesting nodes at different network scales are reviewed as well as models for energy consumption at the nodes.
\end{abstract}

\section{Introduction}

Providing energy harvesting capability to wireless devices enables the nodes to continually acquire energy from nature or man-made phenomena. This in turn provides a promising future for wireless networks: self-sustainability and virtually perpetual operation with network lifetimes limited by those of the hardware rather than the energy storage. {\it Energy harvesting wireless networks} are expected to introduce several transformative changes in wireless networking as we know it: in addition to energy self-sufficiency and perpetual operation, expected benefits include reduced use of conventional energy and accompanying carbon footprint, untethered mobility by breaking away from conventional battery recharging, and an ability to deploy wireless networks at hard-to-reach places such as remote rural areas, within concrete structures, and within the human body. As such, energy harvesting wireless networks will make it possible to develop new medical, environmental, monitoring/surveillance and safety applications which are otherwise impossible with conventional battery-powered operation.

There are several different natural sources and associated technologies for energy harvesting: solar, indoor lighting, vibrational, thermal, biological, chemical, electromagnetic, etc. \cite{IEEEsj:Scavenging, IEEEsj:Microsensor, revisionA, IEEEsj:Roundy_thesis, IEEEsj:PicoRadio, Kansal-1274870, IEEEsj:Microstrain_Harvesting, MFC}. In addition, energy may be harvested from man-made sources via wireless energy transfer, where energy is transferred from one node to another in a controlled manner. These technologies have varying degrees of harvesting capacities and efficiencies. While the devices/circuits side of engineering has been continually working to improve energy harvesting mechanisms and devices and their efficiency, the efforts on the signals/systems side of engineering to develop communication schemes for networks composed of energy harvesting nodes have been quite recent, e.g., \cite{Yates09TWC, tassiulas10TWC, sharma10TWC, tcom-submit, kaya_subm, ozel11, ho2012optimal, jingmac, wless-submit, uysal_paper, finite, elsevier, kaya_subm2, isit-hybrid, it-submit, ozel_asilomar, tutuncuoglu2013binary, hassibi-isit13, gunduz-leakage, gunduz-camsap, Zhang_Relay, luo2012optimal, reddy2011duty, huang2011throughput, orhan_ciss, orhan_sarnoff, orhan_isit, burak_gsip2, relay_kaya, burak_gsip, orhan12ITW, xu2012circuitpower, ElzaarXiv, revision2, vaze2012transmission, tapparello2012dynamic, luo2012training, gregori2011efficient, luo2012diversity, lee2012cognitive, ahmed2012power, castiglione2011energy, reddy2012dual, revisionJ, nayyar2012optimal, blasco2012learning, vaze2011competitive, mitran2012optimal}; see also \cite{Gunduz2014COMMAG}. The goal of this review article is to summarize recent results in energy harvesting wireless communications and wireless energy transfer from the perspectives of communication theory, signal processing, information theory and wireless networking. Energy harvesting brings new dimensions to the wireless communication problem in the form of {\it intermittency} and {\it randomness} of available energy, as well as the possibility of {\it sharing energy} among the nodes in a network via wireless energy transfer, which necessitate a fresh look at wireless communication protocols at the physical, medium access and networking layers, as well as at the fundamental performance limits, i.e., the channel capacity. In this article, we summarize such approaches taken in the past few years in this new research field.

\begin{figure*}[t]
\centerline{\begin{tabular}{cc}
\subfigure{\includegraphics[width=0.45\linewidth]{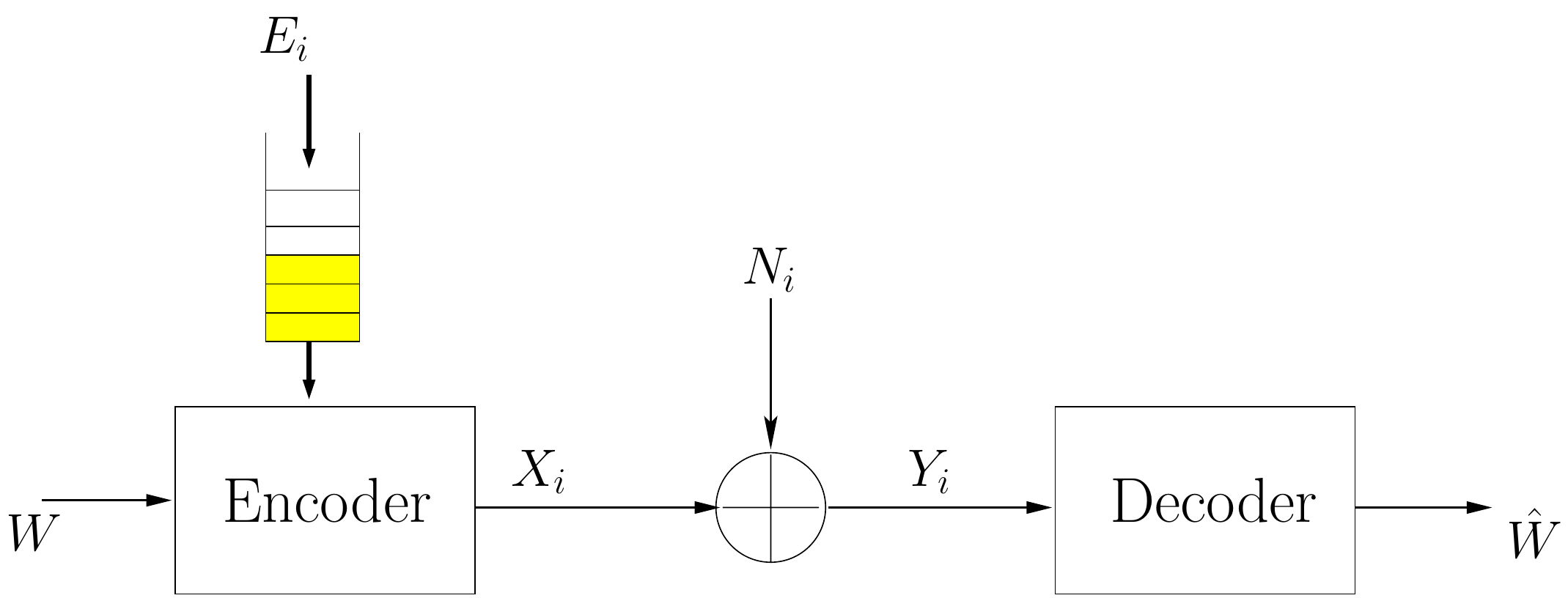}}&
\subfigure{\includegraphics[width=0.45\linewidth]{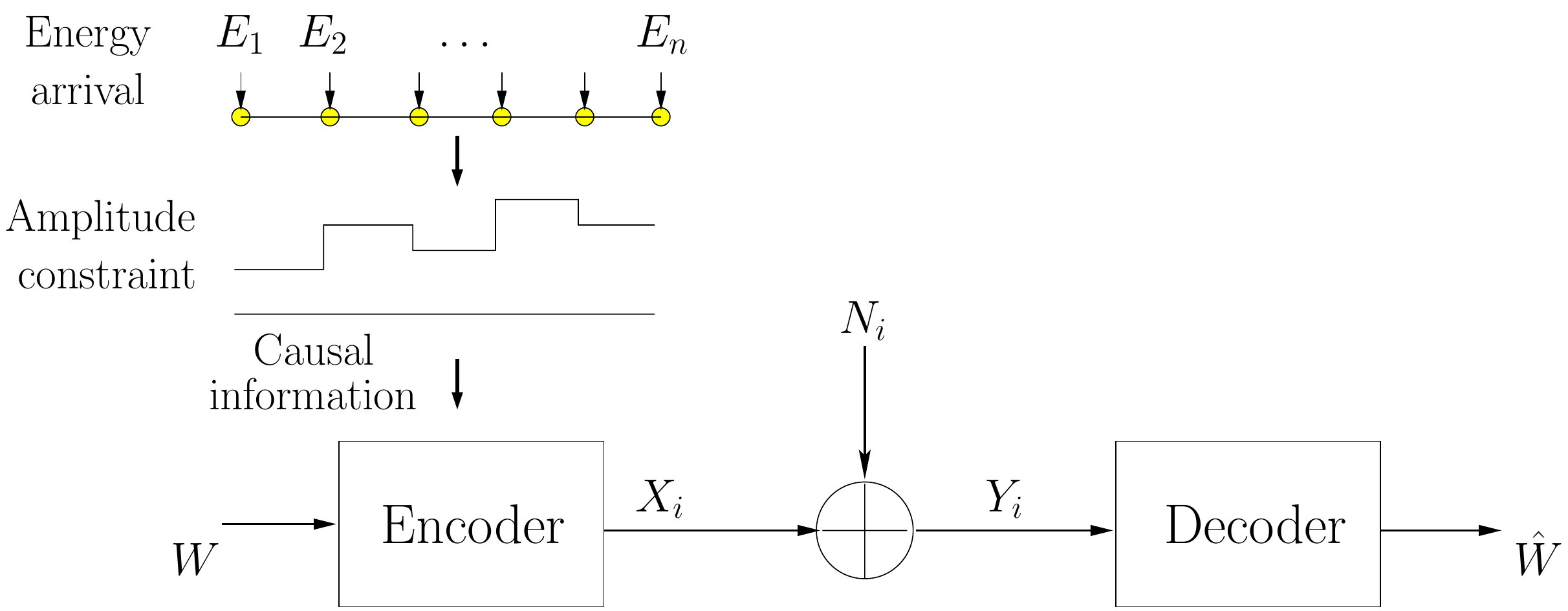}}\\
(a) & (b)
\end{tabular}}
\caption{AWGN channel with random energy arrivals with (a) an unlimited battery, (b) no battery.}
\label{fig1}
\vspace*{-0.3cm}
\end{figure*}

The remainder of the paper is organized as follows. In Section~\ref{eh-it}, we summarize the efforts in identifying the information-theoretic limits of energy harvesting communications, i.e., capacity, by considering energy harvests at the channel use level. Next, in Section~\ref{eh-tm}, we consider throughput maximization, where energy harvests are at the communication slot level, and describe the efforts in identifying the throughput optimal transmission power and scheduling policies. This section considers offline availability of energy harvesting times and amounts. In Section~\ref{new-online-section}, we consider online optimization of general reward functions, including the throughput. In Section~\ref{eh-macprotocols}, we go above the physical layer and develop medium access layer control protocols for energy harvesting wireless devices. Next, Section~\ref{eh-et} considers the case where energy can be shared and transferred between the nodes in a wireless network, jointly with information. Here, we consider both information-theoretic and network-theoretic advances in information and energy transfer including the notions of energy cooperation and interactive exchange of information and energy. In Section~\ref{ehsn}, we consider energy harvesting sensor networks where a number of operations other than transmission incur energy costs and impact the performance. Next, in Section~\ref{lsehn}, we examine the potential of large-scale energy harvesting networks, including mobile ad hoc and cellular networks composed of energy harvesting wireless devices, as well as heterogenous systems with nodes of varying capabilities. Section~\ref{consumption} describes the advances in identifying total energy consumption models for energy harvesting communication systems. Finally, Section~\ref{conclusion} provides the conclusions for the article and lists some future directions in the broad area of energy harvesting communications.

\section{An Information-Theoretic View of Energy Harvesting}
\label{eh-it}
\begin{figure*}[t]
\centerline{\begin{tabular}{cc}
\subfigure{\includegraphics[width=0.45\linewidth]{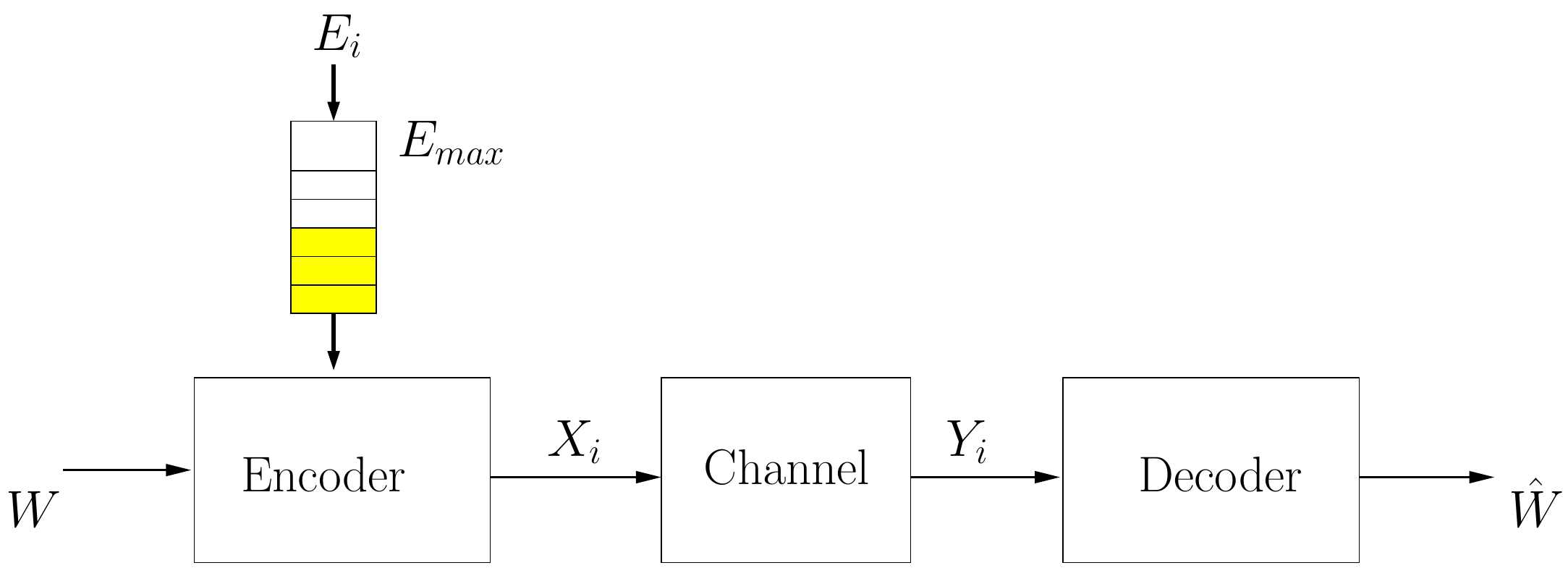}}&
\subfigure{\includegraphics[width=0.45\linewidth]{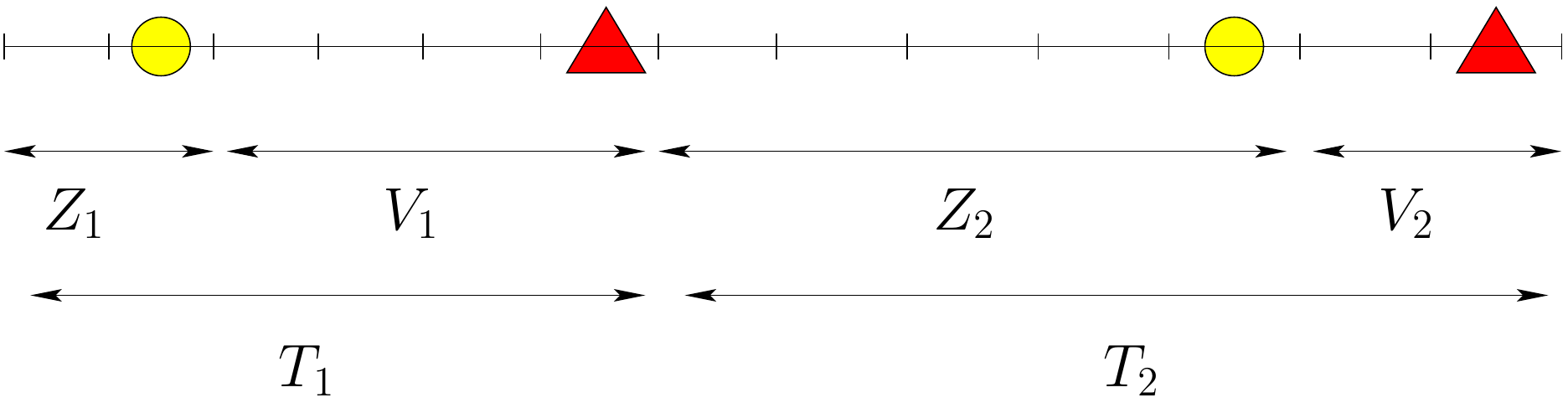}}\\
(a) & (b)
\end{tabular}}
\caption{(a) System with a finite-sized battery, (b) equivalent timing representation.}
\label{fig2}
\vspace*{-0.3cm}
\end{figure*}

Consider the classical AWGN channel with input $X$, additive zero-mean unit-variance Gaussian noise $N$, and output $Y=X+N$. The capacity of this channel is $C = \frac{1}{2} \log(1+P)$. In this classical result of Shannon \cite{shannon1}, the codewords are average power constrained in the following way:
\begin{align}
\frac{1}{n} \sum_{i=1}^n X_i^2 \leq P
\label{avg-pow-cons}
\end{align}
for very large $n$, where $X_i$ denotes the $i$th element of the transmitted codeword. Consider now that the energy arrives (is harvested) stochastically at the transmitter as a stationary and ergodic random process $E_i$, with an average recharge rate $\mathbb{E}[E_i]=P$, as shown in Fig.~\ref{fig1}(a). We therefore introduce a {\it canonical} model of an energy harvesting system as a communication channel augmented with an energy harvesting battery (energy queue) as shown in Fig.~\ref{fig1}(a). In this initial model, we assume that the battery has an unbounded capacity as depicted by an open ceiling in Fig.~\ref{fig1}(a). At each channel use, $X_i^2$ units of energy are depleted from the battery, and $E_i$ units of energy enter the battery. For a codeword to be transmitted without any energy outages, we need  to satisfy the \emph{energy causality constraints} at every channel use
\begin{align}
\label{harv-pow-cons}
\sum_{i=1}^k X_i^2 \leq \sum_{i=1}^k E_i, \qquad k=1,\ldots,n
\end{align}
That is, at each channel use, the cumulative energy expended  cannot exceed the cumulative energy harvested.
We note that while the average power constraint in the classical information theory setting in (\ref{avg-pow-cons}) imposes a single constraint for the entire codeword, the energy harvesting scenario in (\ref{harv-pow-cons}) imposes $n$ power constraints on the codeword. We also note that the cumulative form of the constraints (\ref{harv-pow-cons}), and the unbounded nature of the battery, allow for saving of the energy harvested in any channel use to be used at a later channel use. On the other hand, when there is no battery to save harvested energy for future use, the constraints on the codewords become
\begin{align}
\label{harv-pow-cons-no-battery}
X_i^2 \leq E_i, \qquad i=1,\ldots,n
\end{align}
which impose {\it instantaneous stochastic amplitude constraints} on the code symbols, as depicted in Fig.~\ref{fig1}(b). When we have a finite-sized battery (of maximum size $E_{max}$) as shown in Fig.~\ref{fig2}(a), the battery size (i.e., the amount of available energy in the battery) at channel use $i$, denoted as $B_i$, will evolve as follows
\begin{align}
B_{i+1} = \min\left\{B_i-X_i^2 + E_i, E_{max}\right\}
\end{align}
which denotes that first an $X_i^2$ amount of energy exits the battery (due to the transmission of symbol $X_i$), and then $E_i$ amount of energy is harvested into the battery. Therefore, what is transmitted, i.e., $X_i$, affects the amount of energy in the battery in the next channel use, and how much energy there is in the battery, i.e., $B_i$, affects the allowable set of symbols via the instantaneous amplitude constraint $X_i^2\leq B_i$. Here we note for future reference that: the battery state $B_i$ will be a highly correlated random process over time even when the harvesting process $E_i$ is i.i.d.; actions of the transmitter (i.e., what it sends) affect the future of the battery state; and the transmitter naturally knows the battery state, but the receiver does not.

The capacity of the energy harvesting channel is known only in the cases of unboundedly large battery ($E_{max}=\infty$) \cite{it-submit}, no battery ($E_{max}=0$) \cite{ozel_asilomar}, and for a unit-sized battery ($E_{max}=1$) over a binary noiseless link \cite{tutuncuoglu2013binary}. We will see that, for a Gaussian channel, the channel capacities for $E_{max}=\infty$ and $E_{max}=0$ are very different and they are achieved with vastly different strategies. In particular, when $E_{max}=\infty$, Gaussian codebooks achieve capacity as in the classical setting, whereas when $E_{max}=0$, discrete signalling is optimal. The preliminary result for the case of unit-sized battery over a binary noiseless link in \cite{tutuncuoglu2013binary} shows the richness of and the challenges posed by this problem, and presents an interesting connection to timing channels \cite{anantharam1996bits}. The capacity for general noisy channels with finite capacities (i.e., for any $E_{max}$) is an important open research problem. Some progress has been made in this direction, in terms of developing lower and upper bounds to the capacity, in recent work \cite{revisionD, revisionD1}.

\subsection{Capacity with an Unlimited-sized Battery, $E_{max}=\infty$}

We first note that each codeword satisfying the constraints (\ref{harv-pow-cons}) also satisfies the average power constraint in (\ref{avg-pow-cons}) automatically by the strong law of large numbers as $\frac{1}{n} \sum_{i=1}^n E_i \rightarrow P$. Therefore, the constraints of the energy harvesting system are stricter, and hence the capacity of the energy harvesting system is upper bounded by the classical capacity with an average power constraint equal to the average recharge rate.

Reference \cite{it-submit} gives two schemes to achieve this upper bound. In the first scheme, called {\bf save-and-transmit}, the transmitter saves energy in the first $h(n)$ channel uses (does not transmit anything), and transmits data in the remaining $n-h(n)$ channels uses by using a codebook generated with i.i.d.~Gaussian samples of power equal to the average recharge rate $P$. By letting both $h(n)$ and $n-h(n)$ go to $\infty$, and choosing $h(n)$ as $o(n)$ such as $\log(n)$, we can achieve the AWGN capacity. This can be seen as follows: By saving an infinite amount of energy in the {\it saving phase} we prevent energy outages in the {\it data transmission phase} with probability one, and by choosing $h(n)$ as $o(n)$ we make the rate-hit taken by not transmitting any data during the {\it saving phase} negligible. In the second scheme, called {\bf best-effort-transmit}, we start data transmission right away without a saving period. We construct a Gaussian codebook with average power that is $\epsilon$ (small enough) less than the average recharge rate. At any given channel use, if we have sufficient energy in the battery, we send the corresponding code symbol, otherwise, we send a zero symbol (i.e., not send anything). This creates mismatches between what is in the codebook, and what is actually transmitted, but the number of mismatches remains finite from the strong law of large numbers, and therefore such mismatches are inconsequential with joint typical decoding.

It is important to note that the availability of an unlimited battery is essential in both achievable schemes. In particular, in the save-and-transmit scheme, the unlimited battery enables us to save essentially an unlimited amount of energy in the saving phase to prevent any energy outages in the data transmission phase. In the best-effort-transmit scheme, the unlimited battery enables the energy queue size to blow up sooner or later, preventing any energy shortages after a large enough channel use index, implying that only finitely many mismatches occur. Another important point to note regarding these two schemes is that neither the transmitter nor the receiver need to know the energy arrival process or the current battery energy state. This is again due to the unlimited nature of the battery which smoothes out the randomness in the stochastic energy arrival process, and any battery state information at the transmitter or receiver does not improve the achievable rates. On the other hand, as we will see, in the case of $0\leq E_{max}<\infty$ it is crucial that the transmitter has causal information of the energy arrival to achieve a non-zero reliable rate. However, we note that it is natural that the transmitter has causal knowledge of the energy arrival process, because it observes the incoming energies into its battery.

Finally, we note that \cite{revision1} considers the same problem, and proves the same capacity result in \cite{it-submit} by using a different proof technique which relies on Asymptotically Mean Stationary (AMS) sequences. In addition, references \cite{revisionE, revisionF} extend these results to multiple access channels with energy harvesting transmitters with unlimited-sized batteries; see also \cite[Section~VI]{it-submit}.

\begin{figure*}[t]
\centerline{\begin{tabular}{cc}
\subfigure{\includegraphics[width=0.45\linewidth]{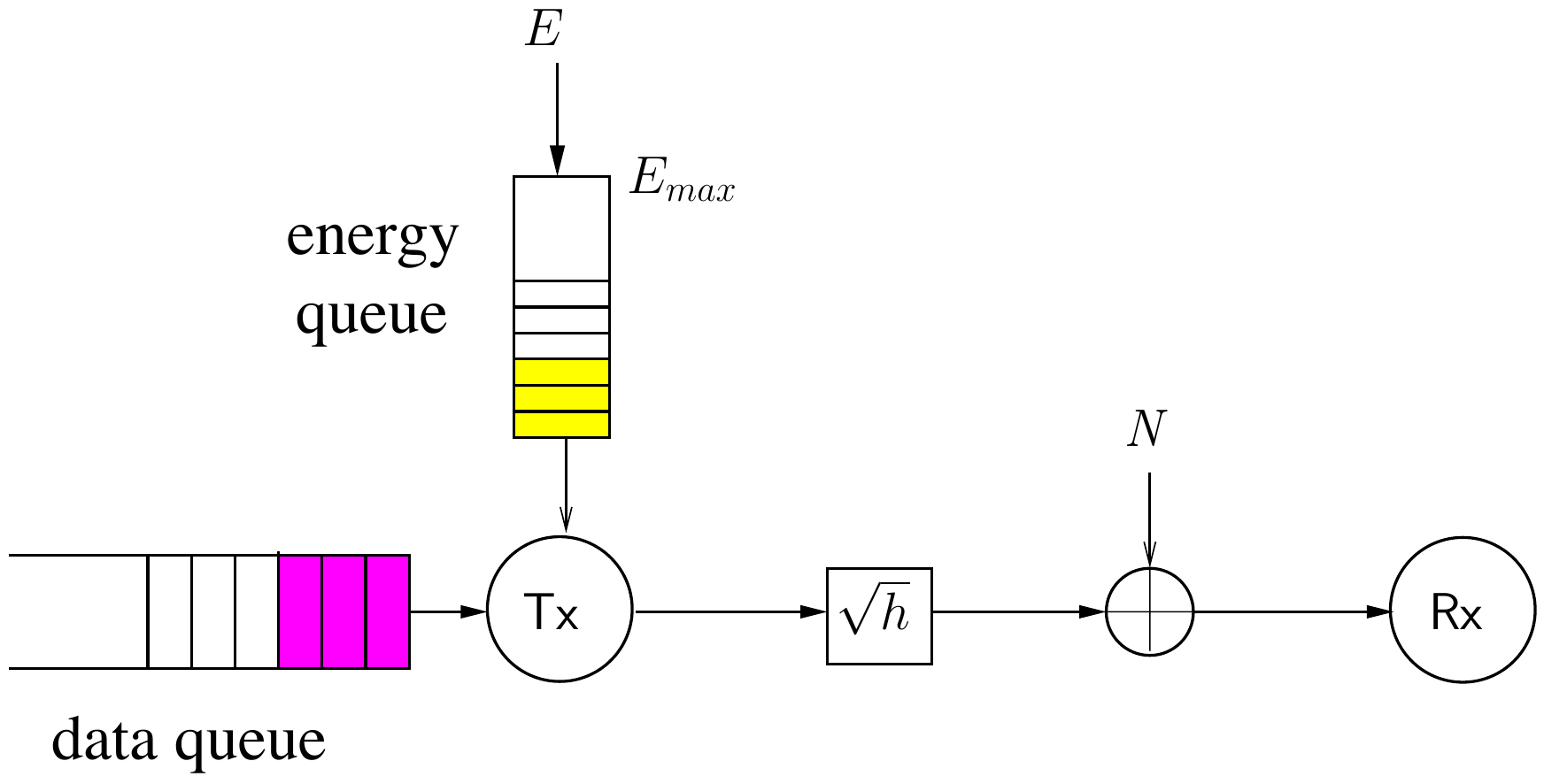}}&
\subfigure{\includegraphics[width=0.35\linewidth]{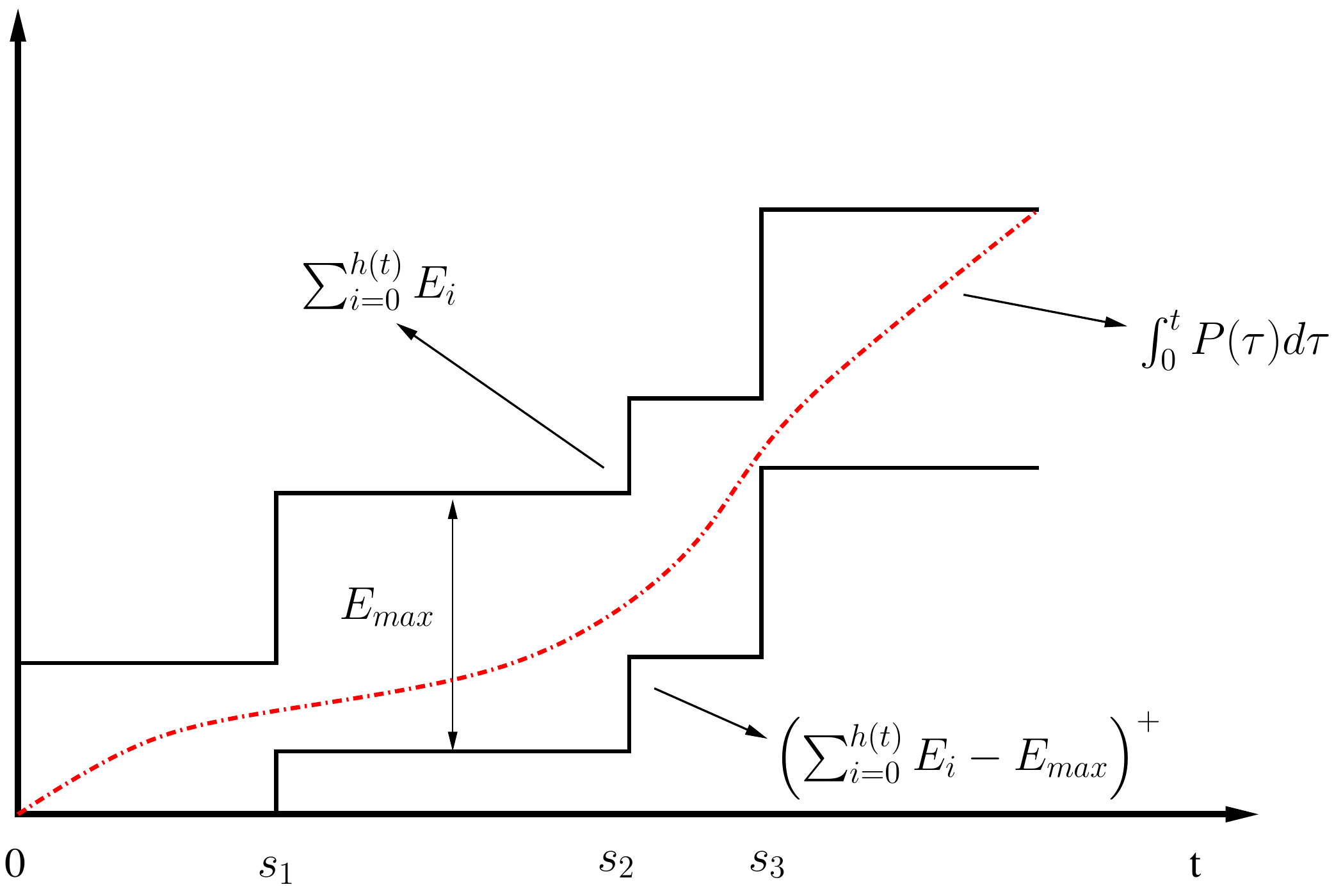}}\\
(a) & (b)
\end{tabular}}
\caption{(a) Energy harvesting system. (b) Energy feasibility tunnel.}
\label{fig3}
\vspace*{-0.3cm}
\end{figure*}

\subsection{Capacity with no Battery, $E_{max}=0$}

We now consider the other extreme where there is no battery to store and save energy for future use. In this case, the channel inputs are instantaneously amplitude constrained as in (\ref{harv-pow-cons-no-battery}). However, different from the existing literature \cite{Smith71}, these amplitude constraints are not deterministic and constant, but are time-varying and stochastic. The transmitter knows the energy arrival profile causally, and the receiver does not know it. In this case, the transmitter can choose the code symbols according to the observed energy, which is the {\it state} of the system. Reference \cite{ozel_asilomar} combines the works of Smith \cite{Smith71}, which considers the static amplitude-constrained AWGN, and Shannon \cite{Shannon58IBM}, which considers the capacity of state-dependent channels with causal state information available at only the transmitter, to determine the capacity of this channel model. In the solution, the transmitter sends the channel input $T_1$ when the energy arrival state is $e_1$ and $T_2$ when the energy arrival state is $e_2$, where $T_1, T_2$ are jointly distributed random variables in $[-\sqrt{e_1}, \sqrt{e_1}] \times [-\sqrt{e_2}, \sqrt{e_2}]$. Reference \cite{ozel_asilomar} shows that the support set of this distribution is of Lebesgue measure zero, i.e., it is discrete. Further, it observes experimentally that the support set of this discrete distribution is finite. This is reminiscent of Smith's result for the static amplitude constrained case, where the input distribution resides in one dimension. In addition, if $e_1$ and $e_2$ are sufficiently small, then symmetric binary distribution with masses at $(\sqrt{e_1},\sqrt{e_2})$ and $(-\sqrt{e_1},-\sqrt{e_2})$ is optimal. As $e_1$ and $e_2$ are increased, the optimal distribution begins to have higher number of mass points, e.g., ternary, quaternary, etc. Reference \cite{ozel_asilomar} shows that the capacity with an unlimited battery is significantly larger than the capacity with no battery. Reference \cite{revisionG} extends these results to the case of an energy harvesting multiple access channel with no batteries.

\subsection{Capacity with Unit-sized Battery, $E_{max}=1$}

We now summarize the very recent work in \cite{tutuncuoglu2013binary} which takes a step towards understanding the capacity of an energy harvesting link with a finite-sized battery, i.e., $0<E_{max}<\infty$; see Fig.~\ref{fig2}(a). With a finite-sized battery, the channel inputs are instantaneously amplitude-limited to the (square root of the) current amount of energy in the battery. From \cite{Smith71, ozel_asilomar} (and the previous sub-section), we know that, when the channel inputs are constant amplitude-constrained, or i.i.d.~stochastic amplitude-constrained, over a Gaussian channel, the optimum input distributions are discrete. However, these discrete mass points are arbitrary real numbers, and it is hard to track the dynamics of the energy queue, if it is served with codebooks generated by arbitrary real mass points. For a tractable abstraction of the system, \cite{tutuncuoglu2013binary} models energy arrivals as multiples of a fixed quantity, and correspondingly, considers a physical layer which has a discrete alphabet based on this fixed quantity. For further analytical tractability, \cite{tutuncuoglu2013binary} assumes that the physical layer is a noiseless binary channel, energy arrivals are binary, and the battery is unit-sized. Even in this simple model, unavailability of the battery state at the receiver, memory of the battery state in time, and the fact that the state evolves based on the previous channel inputs, render the problem challenging. The fact that channel inputs affect future states is reminiscent of action dependent channels in \cite{weissman2010capacity}. While Shannon strategy, which is optimal in the zero-battery case in \cite{ozel_asilomar}, yields achievable rates for the finite-battery case, the transmitter may utilize the memory in the battery state to achieve higher rates.

Reference \cite{tutuncuoglu2013binary} shows that this noiseless binary channel with a unit-battery can equivalently be modeled as a timing channel \cite{anantharam1996bits}, where information is transmitted by timings between 1s, as opposed to the actual places of 1s and 0s. This converts the problem into a timing channel with additive geometric noise (service time), where the service time is causally known to the transmitter. This is explained in Fig.~\ref{fig2}(b), where circles represent energies harvested and triangles represents 1s put to the channel. Here, after sending a 1, we have to wait a random $Z_i$ number of channel uses to receive an energy into the battery. Then, we choose to wait a $V_i$ number of channel uses to send information. Here $V_i$ serves as our channel input that is to be optimized. Reference \cite{tutuncuoglu2013binary} combines Anantharam-Verdu's bits through queues \cite{anantharam1996bits} and Shannon's state-dependent channels with causal state information available at only the transmitter \cite{Shannon58IBM} to find a single-letter capacity expression for the capacity of this equivalent channel. We note that the channel input $V_i$ is a function of the message (denoted by $U_i$ in \cite{tutuncuoglu2013binary}) and the state $Z_i$ which is causally known to the transmitter. Shannon strategy \cite{Shannon58IBM} is optimal here, because the state $Z_i$ is i.i.d.~in time. The capacity expression involves an auxiliary random variable $U$, and its optimization is difficult. For this reason, reference \cite{tutuncuoglu2013binary} determines an achievable rate based on a certain selection of this auxiliary random variable. This selection resembles the concentration idea in \cite{willems2000signaling}, and may be interpreted as a lattice-type coding for the timing channel. Reference \cite{hassibi-isit13} provides an $n$-letter expression for the noisy-channel version of this problem with an arbitrary battery size, conjectures that it is the capacity, and evaluates it using techniques in \cite{arnold06}.

\section{Offline Energy Management for Throughput Maximization}
\label{eh-tm}
In this section, we take a communication theory and networking approach to the energy harvesting communication problem. We first consider the basic single-user channel, and then present extensions to multi-user settings and practical considerations such as processing costs and battery imperfections.

\begin{figure*}[t]
\centerline{\begin{tabular}{cc}
\subfigure{\includegraphics[width=0.45\linewidth]{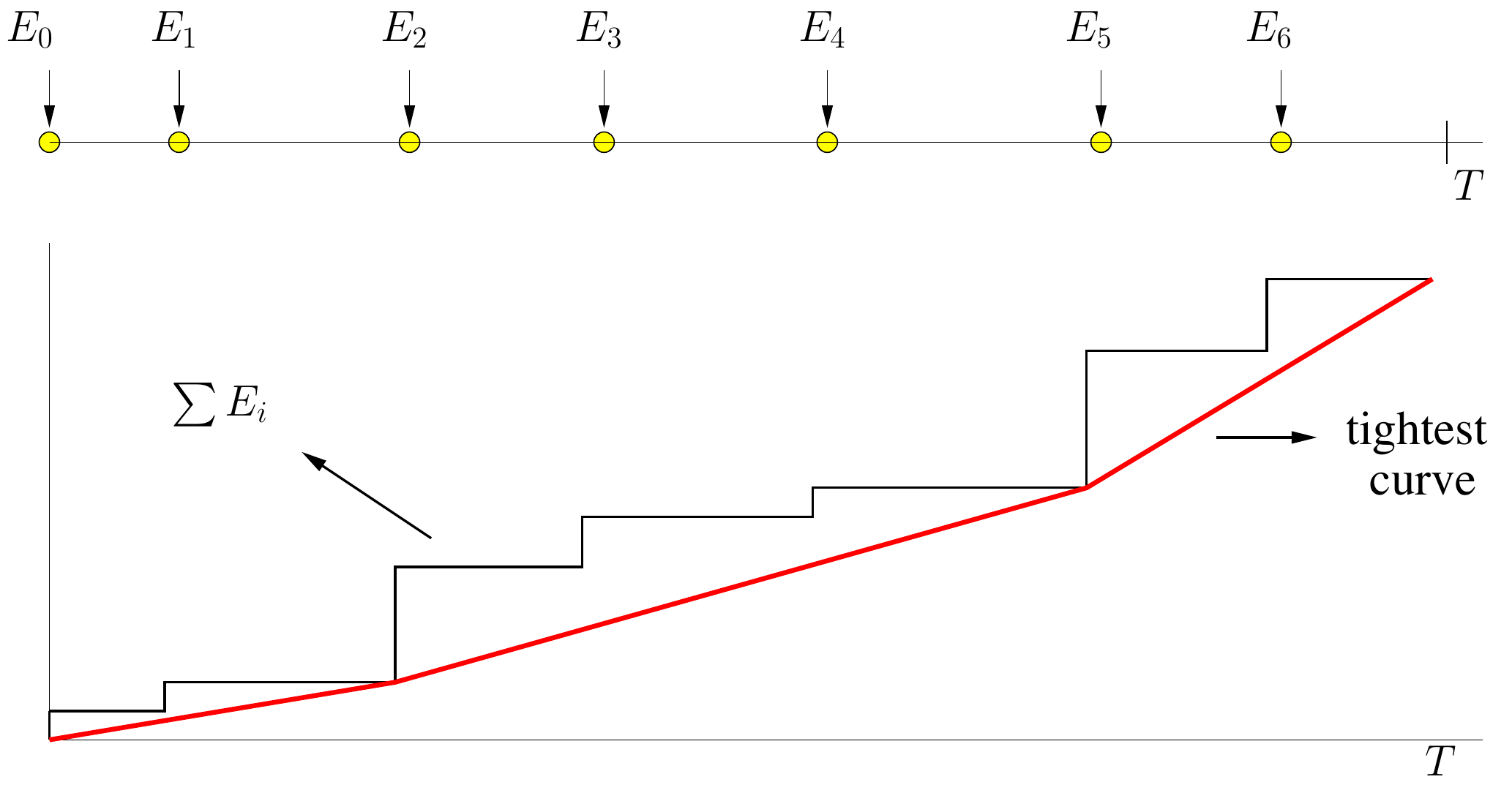}}&
\subfigure{\includegraphics[width=0.45\linewidth]{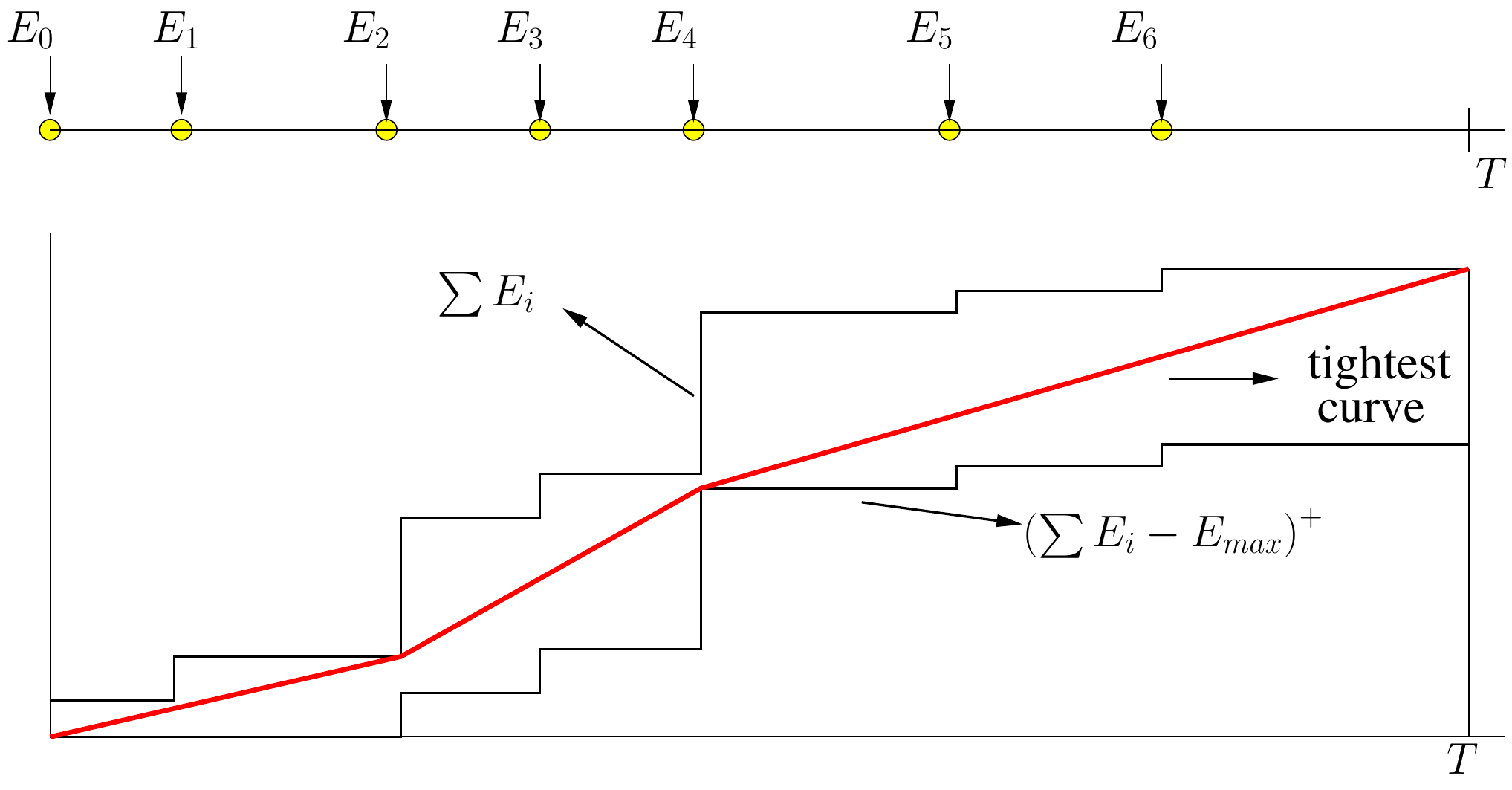}}\\
(a) & (b)
\end{tabular}}
\caption{The optimal policy is the tightest string that connects the two ends. (a) $E_{max}=\infty$. (b) Finite $E_{max}$.}
\label{fig4}
\vspace*{-0.3cm}
\end{figure*}

\subsection{Single-User Channel}

Consider the single-user fading channel with additive Gaussian noise as shown in Fig.~\ref{fig3}(a). The transmitter has two queues, the data queue where the data packets are stored, and an energy queue where the arriving (harvested) energy is stored. The goal here is to schedule the transmission of data packets in the data queue using the energy in the battery. We relate the instantaneous power and rate through a monotone increasing concave function. While we can use an arbitrary monotone concave relationship, for simplicity and convenience, we assume the following familiar power-rate relationship: $R=\frac{1}{2}\log(1 + h P)$. Therefore, whenever we send a signal with power $P$ in an epoch of duration $\ell$, $\frac{\ell}{2} \log\left(1 + hP\right)$ bits of data are served out from the data backlog with the cost of $\ell P$ units of energy depletion from the energy queue. With this model in mind, we solve for the optimum power control policy $P(t)$ in time as a function of the energy arrival profile, the data backlog profile and the channel fading profile, in order to minimize the time by which all of the packets are successfully transmitted. Minimizing the transmission completion time for a given number of bits is equivalent to maximizing the number of bits transmitted in a given duration. Therefore, in the following, we consider maximizing the number of bits delivered by a deadline $T$.

The optimization problem is subject to the {\it energy causality} constraint on the harvested energy, and the {\it finite-storage} constraint on the rechargeable battery. In particular, the {\it energy causality} constraint requires that the energy that has not arrived yet (has not been harvested yet) cannot be used. The {\it finite-storage} constraint, on the other hand, requires that no energy is wasted because of battery being full at the time of energy arrivals; we also call this constraint the {\it no-energy-overflow} constraint. Assume that energies of $\{E_0, E_1, \ldots, E_{N-1} \}$ are harvested, and epoch lengths are $\{\ell_1,\ldots,\ell_{N-1} \}$. Due to the concavity of the rate-power relationship, power must be kept constant between energy harvests \cite{tcom-submit}. This reduces the power control policy of $P(t)$ to a sequence of constant powers $\{p_1,\ldots,p_N\}$. The energy causality constraints become \cite{tcom-submit}
\begin{align}
\label{const1}
\sum_{i=1}^{k} \ell_i p_i \leq \sum_{i=0}^{k-1} E_i , \quad k=1,\ldots, N
\end{align}
and the no-energy-overflow constraints become \cite{kaya_subm}
\begin{align} \label{const2}
\sum_{i=0}^{k} E_i - \sum_{i=1}^{k} \ell_ip_i \leq E_{max}, \quad k=1,\ldots,N-1
\end{align}

We illustrate these two constraints on the energy consumption policy in Fig.~\ref{fig3}(b). The upper {\it staircase} is the cumulative energy arrival profile which provides the {\it energy causality} upper bound, and the lower {\it staircase} is the {\it no-energy-overflow} curve which provides a lower bound. Any feasible energy consumption curve must lie in between. We note that the {\it energy causality} constraint forces the energy consumption to slow down not to exceed the harvested amount, while the {\it no-energy-overflow} constraint forces energy consumption to speed up to open up space in the battery for new energy arrivals. Although the optimization is over all monotonically non-decreasing time functions for the {\it energy consumption curve,} by the concavity of the objective function, the optimal policy must remain constant in between energy harvests; therefore, the dimension of the optimization problem is reduced to the finite number of epochs in an interval. Geometrically, this means that the feasible energy consumption profiles which are candidates to be optimum must be piece-wise linear. The optimal policy is shown to be the {\it tightest string} that lies in the {\it energy feasibility tunnel} \cite{tcom-submit, kaya_subm}. This solution aims to {\it keep longest stretches of constant power periods} subject to energy causality and no-energy-overflow constraints, as the concavity of the power-rate relationship favors constant powers to the extent possible. An example of the optimum energy consumption curve is shown in Fig.~\ref{fig4}(a) for $E_{max}=\infty$, i.e., there is no energy overflow concerns, and in Fig.~\ref{fig4}(b) for a finite $E_{max}$.

\begin{figure*}[t]
\centerline{\begin{tabular}{cc}
\subfigure{\includegraphics[width=0.45\linewidth]{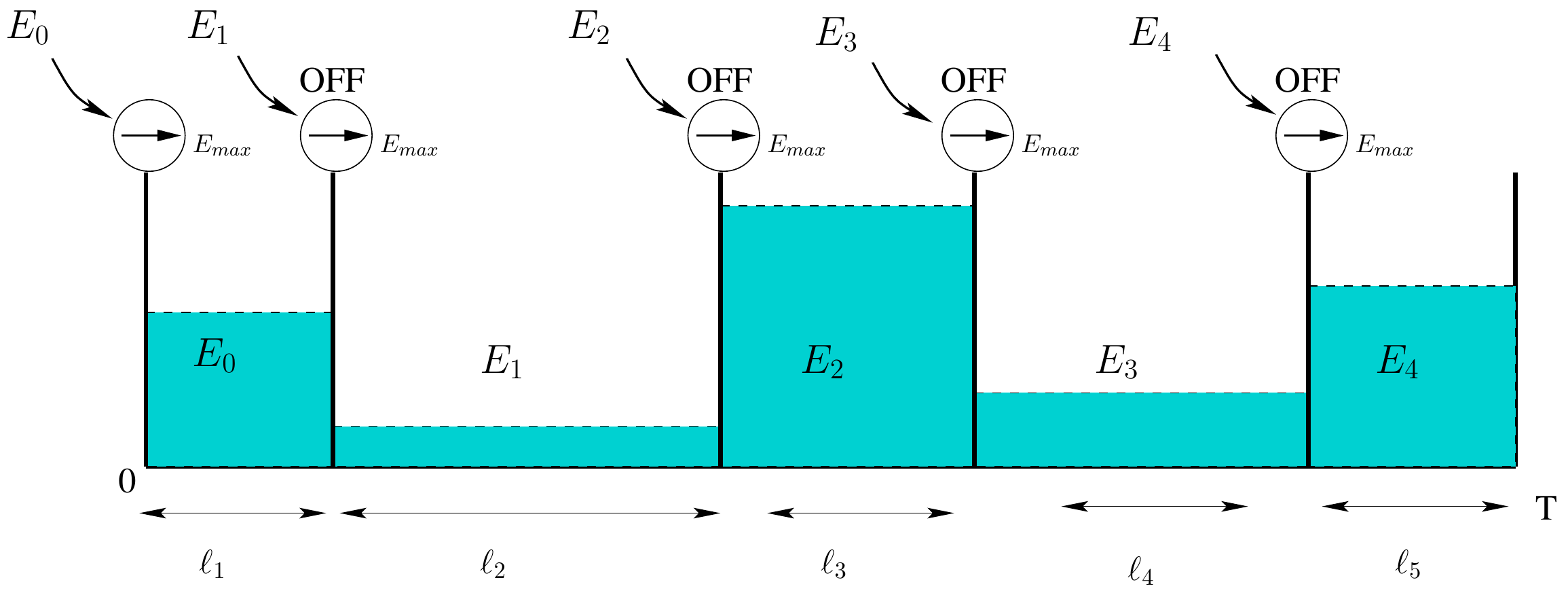}}&
\subfigure{\includegraphics[width=0.45\linewidth]{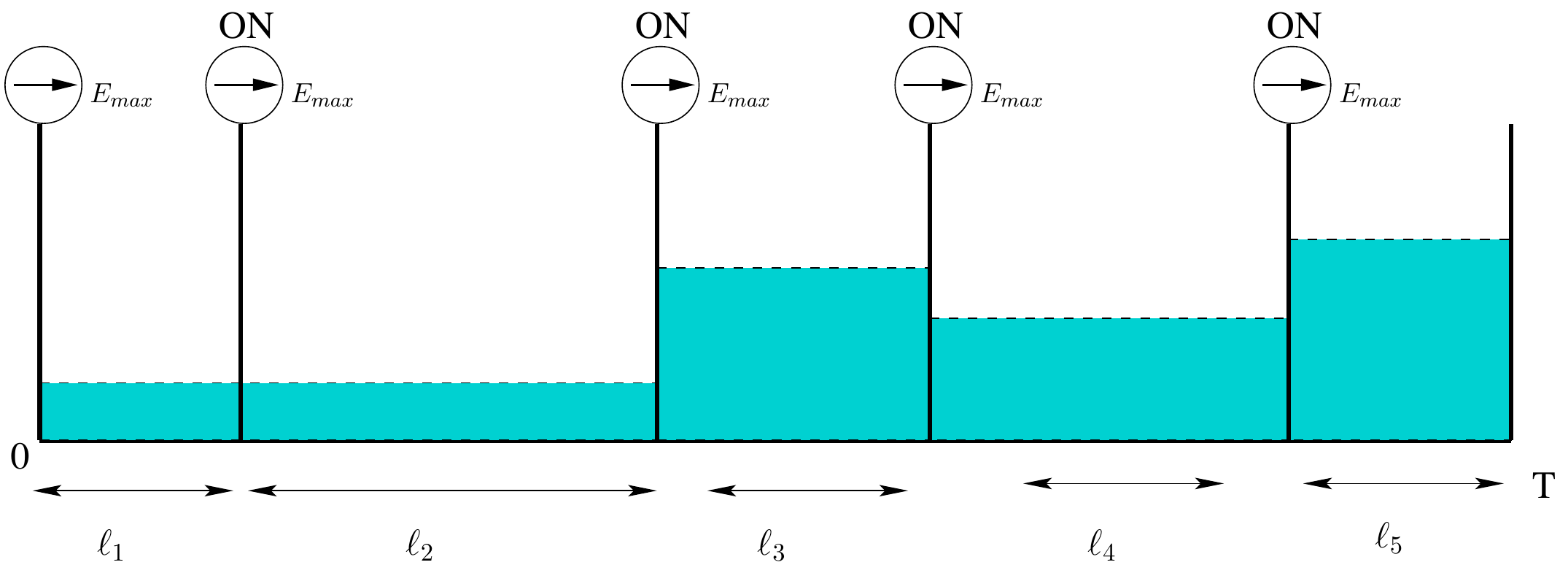}}\\
(a) & (b)
\end{tabular}}
\caption{Directional water-filling algorithm. (a) Initial water (energy) levels. (b) Final water levels.}
\label{fig5}
\vspace*{-0.3cm}
\end{figure*}

An alternative approach to the {\it feasible tunnel} approach is the {\it directional water-filling} algorithm presented in \cite{ozel11}. The directional water-filling algorithm aims to distribute the water (energy) equally over time, subject to {\it energy causality} constraints, which introduce the {\it directionality} of water (energy) flow. The directional water-filling algorithm requires walls at the points of energy arrival, with right permeable water taps in each wall which allows water to flow only to the right. This implements the {\it energy causality} constraint, i.e., energy can be saved and used in the future, but the energy that will arrive in the future cannot be used before it has arrived. In addition, these taps allow at most $E_{max}$ amount of water to flow to the right. This implements the {\it finite-capacity battery} constraint by avoiding overflows. These are based on the KKT optimality conditions found from the corresponding convex optimization problem \cite{ozel11}
\begin{align}
\label{ss1}
p_i^* = \frac{1}{\left(\sum_{j=i}^{N+1} \lambda_j - \sum_{j=i}^{N} \mu_j  \right)}-1, \quad i=1,\ldots,N
\end{align}
where $\lambda_i$ are the Lagrange multiplier that enforce energy causality and $\mu_i$ are the Lagrange multipliers that enforce no-energy-overflow conditions. In the implementation of the directional water-filling algorithm, first, the taps are kept off, and transfer from one epoch to the other is not allowed. Then, the taps are turned on one by one, and at most $E_{max}-E_i$ units of energy transfer from past to the $i+1$st epoch is allowed. An example run of the algorithm is shown in Fig.~\ref{fig5}, for a case of 5 epochs. Four energy arrivals occur during the course of the transmission, in addition to the energy available at time $t=0$. We observe that the energy level equalizes in epochs 1 and 2. The energy arriving at the beginning of epoch 3 cannot flow left due to the energy causality constraint, which are enforced by right permeable taps. We observe that the excess energy in epoch 3 cannot flow right either, due to the $E_{max}$ constraint. In addition, the energy arriving at the beginning of epoch 5 cannot flow left, again due to the energy causality constraint. In the case of a fading channel, additionally, the strengths of the channel states play an important role in the directional water-filling algorithm \cite{ozel11}. See also \cite{ho2012optimal} for a {\it staircase water-filling} algorithm.

\subsection{Multi-User Channels and Practical Considerations}

This concludes the summary of the basic findings in the case of a {\it single-user} channel. In the case of a {\it broadcast channel} with an energy harvesting transmitter, references \cite{wless-submit, finite} showed that the optimal {\it total} transmit power management policy is the same as the optimal single-user counterpart summarized above, and this optimum total transmit power is distributed among signals going to the users  according to a {\it cut-off} structure; only the optimum total transmit power that is above this cut-off level goes to the weaker user. For the case of a broadcast channel, an iterative algorithm is developed in \cite{uysal_paper}, and these approaches are generalized to a fading and MIMO case in \cite{elsevier}. In the case of a {\it multiple access channel} with energy harvesting transmitters, reference \cite{jingmac} uses a combination of directional water-filling together with a generalized water-filling in \cite{kaya_ulukus_tw1} and iterative water-filling in \cite{weiyu} to obtain the optimum energy management schemes to maximize the region of departed bits in a given duration. For the case of an {\it interference channel}, using some recent advances in the sum-capacity of the interference channel, reference \cite{kaya_subm2} develops sum-rate optimal transmission policies for a class of two-user interference channels with energy harvesting transmitters. In particular, using concavity properties of the sum-rate expressions, sum-rate optimal transmission policies of the users are found by using directional iterative water-filling algorithms, where one user's power profile determines the noise profile of the other user.

References \cite{Zhang_Relay,gunduz-camsap} consider the full duplex end-to-end communication over a two-hop {\it relay channel} where the source and the relay node harvest energy from nature. The source sends data by using its harvested energy and the relay forwards the data coming from the source to the destination using its own harvested energy. References \cite{Zhang_Relay, gunduz-camsap} show that the optimal policy is in general non-unique and that there exists a separable optimal policy in which the source optimizes its throughput without regard to the relay energy harvesting profile, and the relay optimizes its throughput subject to its own energy profile and the data profile coming from the source. In \cite{orhan_ciss, orhan_sarnoff, orhan_isit, burak_gsip2, relay_kaya, burak_gsip}, half-duplex two-hop and more general two-way relay settings are studied.

Reference \cite{orhan12ITW, xu2012circuitpower} consider the case of {\it processing costs}, where the transmitter spends a constant amount of energy per unit time for the circuitry when the transmit power is non-zero, and solve the throughput maximization problem subject to energy causality, no-energy-overflow and processing cost conditions. The solution is characterized as a {\it directional glue-pouring algorithm}. There is a threshold power level $p^{*}$ that is found by solving the following fixed-point equation:
\begin{align}
\frac{\log(1+p^{*})}{p^{*} + \epsilon} = \frac{1}{1+p^{*}}
\end{align}
Glue-pouring is performed such that the power level is always higher than $p^{*}$ whenever it is non-zero and the glue level is calculated accordingly. In particular, the optimal transmission policy is bursty in the sense that the length of a transmission schedule is not allowed to be arbitrarily long due to the processing cost incurred per unit time. In the directional glue-pouring algorithm, harvested energies are allocated into the corresponding epochs first where energy is viewed as the glue in \cite{Glue2008}. Then, the glue is allowed to flow to the right only and the equilibrium glue levels are determined. Reference \cite{ElzaarXiv} extends the results in \cite{orhan12ITW} by considering a broadband fading energy harvesting communication system with processing cost, and shows that when energy is limited, additional bandwidth may not be utilized.

References \cite{gunduz-leakage, kaya_subm2, isit-hybrid} address practical imperfections in energy storage and retrieval. Energy storage units may foster imperfections such as losses during charging/discharging, leakage of available energy over time, etc. In particular, \cite{gunduz-leakage} considers offline throughput maximization for energy harvesting systems with {\it energy leakage over time}. The presence of energy leakage is reflected in the modification of the shape of the energy feasibility tunnel. In particular, the cumulative energy harvesting curve decreases in between two energy harvests due to energy leakage and the power levels have to be determined accordingly. Reference \cite{kaya_subm2} considers the imperfection that occurs instantaneously at the time of charging and discharging of the battery. In particular, due to the {\it inefficiency in charging/discharging}, \cite{kaya_subm2} shows that it may be more advantageous to immediately use the harvested energy without first storing it in the battery, and determines an optimum double-threshold policy. Recent reference \cite{isit-hybrid} considers the case where the portion of the energy not stored in the inefficient battery can be saved in an efficient, but size-limited, super-capacitor, yielding a {\it hybrid energy storage} unit composed of a large-sized but inefficient battery and small-sized but perfect super-capacitor. Reference \cite{isit-hybrid} finds the optimum power management policy by applying the directional water-filling algorithm in multiple stages. Reference \cite{revision2} considers the outage probability as opposed to throughput as the design metric, and determines the corresponding optimal power allocation policy.

\section{Online Energy Management for General Reward Maximization} \label{new-online-section}

In this section, we consider the case in which devices with energy harvesting capabilities send data packets to a receiver according to some transmission policy. Specifically, time is slotted, and in every slot a data packet is produced, which can be either transmitted in the same slot or discarded (i.e., no queueing is considered). The transmission of a packet corresponds to some reward, and maximization of the long-term average reward per slot is sought. We start by considering the case with a single device. The treatment here is based on references \cite{IannelloCISS, MichelusiTCOM}. Related work that studies the throughput of TDMA and carrier sense multiple access protocols can be found in \cite{revisionK}.

The device operates in a completely on-line fashion, i.e., it has only causal knowledge of the evolution of the system. In this case, the device must make intelligent decisions about whether or not to transmit based on the system state, which includes the amount of energy stored in the battery, and possibly some state of the harvesting process. A useful framework to study these types of problems is provided by Markov Decision Processes. For simplicity, assume that the harvesting process is i.i.d.,\footnote{Extensions to non-i.i.d. harvesting are discussed in \cite{MichelusiTCOM}} so that the state of the system is limited to the
current contents of the battery. Each system state is assigned a set of possible actions (e.g., idle and transmit, possibly with different power levels), so that transitions from a certain state depend on external events (e.g., whether or not some energy is harvested) as well as on the decision made (e.g., whether or not the packet is transmitted). Since each policy (i.e., the definition of the probability distributions of the various actions as a function of the state) will result in a different evolution of the underlying Markov chain and correspondingly a different value of the overall reward, the objective is to come up with the policy for which the reward is maximized.

A quite general model that describes the evolution of the battery status in slot $k$, $B_{k}$, is $B_{k+1}=\textrm{min}\{[B_{k}-Q_{k}]^++E_{k},E_{max}\}$, where $Q_{k}$ is the amount of energy used in slot $k$ as a result of the action chosen, $E_{k}$ is the amount of energy harvested during slot $k$, and $E_{max}$ is the finite storage capacity of the battery.

\subsection{Optimal Transmission Policies with Perfect Knowledge of the State-of-Charge}

Consider the case in which the device has full access to the information
regarding the exact energy level in its own battery. In this case, the battery state evolution above can be observed, so that the decision will always be compatible with the amount of energy that is available in the battery.

Two different models have been studied in this case. In the first model \cite{MichelusiITA12}, each packet has an associated random importance value, $V_{k}$, which represents the reward that is gained if the packet is transmitted. In this case, the action is binary and is represented by the decision about whether or not to transmit the packet. Such decision can be implemented by a simple threshold policy, where the packet is sent if its importance value is above a certain state-dependent threshold, and discarded otherwise. The long-term average reward per slot can be computed as
\begin{equation}
R=\underset{K\rightarrow\infty}{\lim\inf}\frac{1}{K}E\left[\sum_{k=0}^{K-1}Q_{k}V_{k}\right]
\end{equation}
and the optimal strategy can be numerically found using standard techniques, such as the Policy Iteration Algorithm (PIA) \cite{Bertsekas2005}. In some cases, it is possible to approximate the optimal policy through
some heuristics which, while being very easy to compute and to store in the device, provide a level of performance that is very close to the optimum. A notable example, which is asymptotically optimal for
$E_{max}\to\infty$ and very good already for quite modest values of $E_{max}$, is the so-called \emph{balanced policy}, where the threshold is such that in each slot the average consumed energy is equal to the expected harvested energy.

In the second model, different power levels can be used for transmission, which correspond to different rewards (no importance value needs to be considered in this case). The optimization now looks for the best strategy in selecting the transmit power so as to make the best use of the available energy while accounting for the specific relationship between transmit power and reward gained.

As discussed in \cite{MichelusiTCOM}, the above model can be extended to the case in which the harvesting process is correlated, which in most cases is a much more accurate model of what may happen in reality (e.g., solar energy). In this case, if the harvesting process is itself regulated according to some underlying Markov chain, we can include it in the system state and use the same approach as before. Note that in this case the standard numerical procedures can still be applied (at least as long as the complexity of the model is manageable), but the additional complexity may make it more difficult (and in some cases impossible) to obtain closed-form results. An interesting observation in this case is that the performance is no longer determined by the battery size alone, $E_{max}$, but rather depends on the ratio between $E_{max}$ and the dynamics of the harvesting process (e.g., for slow harvesting processes in which periods of low harvest can potentially last long, a large battery is needed for good performance).

\subsection{Optimal Transmission Policies with Imperfect Knowledge of the State-of-Charge}

Although it may seem quite natural that a microprocessor is able to access the battery and to accurately know the amount of energy it contains, several studies have shown that this may not always be the case. For example, uncertainty in the parameters of the components of the circuitry may lead to errors as large as 30\% \cite{Chiasson2005}. In addition, there is a trade-off between the accuracy in the knowledge of the battery State-of-Charge and the amount of time and energy invested in gaining such knowledge, so that, even if it were possible, obtaining accurate information may be too expensive, especially in simple and resource constrained devices. It is therefore of interest to study the case in which the optimization of the transmission policy needs to be carried out under the constraint of uncertain information about the battery status.

The underlying process that describes the evolution of the battery status is still a Markov chain, but in this case the state of the chain is known with some error, and therefore the Markov Decision Process framework no longer applies. Rather, this new formulation falls in the realm of the so-called Partially Observable MDPs, which model the lack of perfect observability of the system state. In this case, the optimal policy can be found, but full knowledge of the past history is required (unlike for the Markov case, where the most recent state is sufficient), making the policy very difficult to compute in general. If resource constrained devices are considered, it may make sense to study simplified policies, in which for example only the latest state is used. For some cases of interest, it has been shown that such suboptimal policies actually provide very good results, and may be very close to the true optimum \cite{Michelusi2012}. For the case of correlated harvesting, it has been shown that overall it is more important to know the harvesting state rather than the energy state \cite{IWCMC2012}.

\subsection{Optimal Transmission Policies for Two Users}

As a first step towards multi-user energy harvesting systems, consider a small network with only two devices who send their data to a Fusion Center (FC) through a collision channel, where individual transmissions are correctly received whereas simultaneous transmissions lead to the loss of both packets. As an upper bound to the performance of such a system, consider the case in which a central controller (e.g., the FC itself) has full
knowledge of the energy state and of the importance values of the packets at the two devices, and decides which (if any) is going to transmit. This case can be studied by extending the previous formulation to a two-dimensional model, which is conceptually similar to the one-dimensional case (and can similarly be solved numerically using the PIA), but does not lend itself to analysis and closed-form results. What can be numerically observed in this case is that as long as the batteries are not very small, a balanced policy is able to provide good results which lose very little compared to the true optimal solution \cite{Testa2013}.

\subsection{Battery Degradation Phenomena}

In traditional energy conservation schemes, it is assumed that a finite amount of energy is initially stored in a non-rechargeable battery, and therefore needs to be judiciously used in order to maximize the device lifetime or some measure of energy efficiency. Energy harvesting technologies and rechargeable batteries hold the promise of breaking this barrier, enabling perpetual operation if the consumption and harvesting of energy can be balanced in some ways. However, this vision overlooks an important aspect, which is familiar to all cell phone users: after a battery has been recharged many times, it starts losing its capability of storing energy and/or it discharges more quickly. In fact, a battery cannot last forever, and some aging phenomena make it less and less capable as a function of the number of recharge cycles it went through. Perpetual operation may not be possible, after all.

In order to study this phenomenon, one can still use the MDP model described above, but needs to introduce an additional layer of memory which tracks the degradation of the battery, which in this case translates in a time-varying characterization of its parameters. For example, after a certain number of cycles, $c$, a battery that originally was capable of storing an amount of energy equal to $E_{max}$ may exhibit a reduced capacity $\alpha(c)E_{max}$, where $\alpha(c)\leq 1$ is a non-increasing function of $c$. In order to avoid the very large number of states that would result from this direct formulation, one can characterize this degradation process using some probabilistic technique, in which several degradation stages are identified and the battery jumps from one to the next with some (small) probability in every slot. Such formulation, described in detail in \cite{TCOM2013_batt}, makes it possible to carefully study these effects while keeping the model complexity within reasonable limits. Using this model, one can characterize the total amount of energy that a battery can provide through all the recharging cycles it can support during its entire life.

\subsection{Sensing Policies}

A formulation that is similar to the above but has a different flavor is presented in \cite{Jaggi2007}. Here the model focuses on the energy consumed for sensing (instead of transmitting), and the policy makes a decision about whether a sensor should be activated in a given slot (with the corresponding energy consumption and accrued benefit) or it should remain idle (to save energy but at the risk of missing important events).

For correlated events, the optimal strategy leverages on the ability to predict whether or not an event is likely to occur in the next slot, and on some thresholding rule on the amount of available energy: intuitively, if an event is expected and there is enough energy, the sensor will be activated.

\begin{figure*}[t]
\centerline{\includegraphics[width=0.8\linewidth]{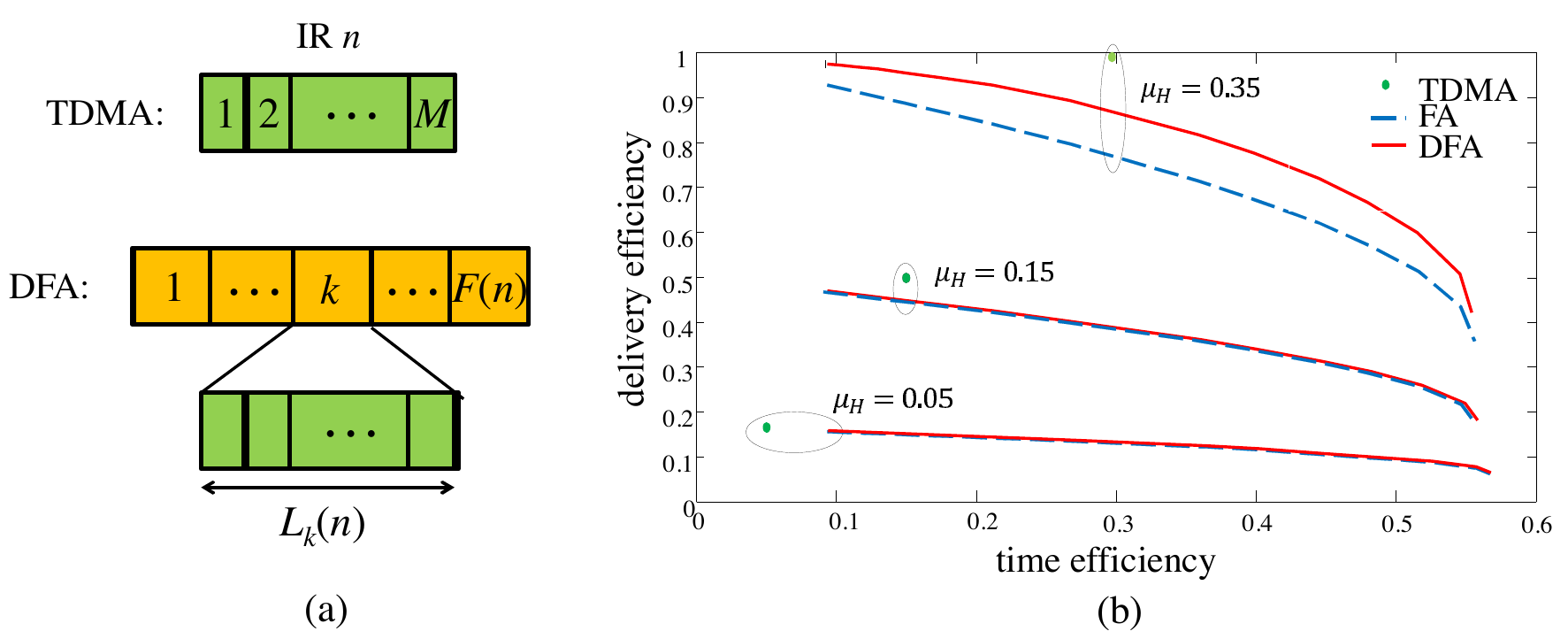}}
\caption{(a) Frames (orange) and slots (green) for TDMA and DFA MAC protocols
in the $n$th inventory round (IR); (b) Trade-off between delivery
and time efficiencies for different harvesting rates $\mu_{H}$ for
the MAC protocols TDMA, FA and DFA.}
\label{fig:efficiencies_region_fixed_gamma}
\vspace*{-0.3cm}
\end{figure*}

\subsection{Optimal Random Multi-Access for a Network of Energy Harvesting Devices}

We now turn to the case in which multiple devices try to access the channel randomly in order to send a data packet of random importance to the Fusion Center (FC). The scheme is very simple in that there is no coordination enforced by the FC. The model for each device would still obey the recursive formulation for the battery state above. The problem here, unlike in the simple two-user example given above, is that no coordination is available, which makes the model much more realistic but also prone to collision errors and inefficiencies.

The reward model for a given user $u$ can be rewritten as
\begin{equation}
R^{(u)}=\underset{K\rightarrow\infty}{\lim\inf}\frac{1}{K}E\left[\sum_{k=0}^{K-1}Q_{u,k}V_{u,k}\prod_{i\neq u}(1-Q_{i,k})\right]
\label{eq:multiuser}
\end{equation}
where the user index has been explicitly added to the decision variable, $Q_{u,k}\in\{0,1\}$, and to the importance value, $V_{u,k}$, and $U$ is the total number of users. Equation (\ref{eq:multiuser}) shows that user $u$ gains its importance value if it decides to transmit while all other users are silent. The overall system reward is therefore expressed as the sum of $R^{(u)}$ over all users $u$. Using the Markov formulation, it is possible to write the objective function as
\begin{equation}
R=\sum_{u=1}^{U}G(\eta_{u})\prod_{i\neq u}(1-P(\eta_{i}))
\label{eq:R}
\end{equation}
where $\eta_{u}$ is the policy for user $u$, which specifies for each level of battery energy, $e=0,1,\ldots,E_{max}$, the transmission probability or, equivalently, the threshold on the importance value used in making a decision about whether or not to transmit, and $G(\eta_{u})$ and $P(\eta_{i})$ are the corresponding average gain and average transmit probability for users $u$ and $i$, respectively. If we assume a symmetric system with fairness constraints, (\ref{eq:R}) becomes
\begin{equation}
R=UG(\eta)(1-P(\eta))^{U-1}
\end{equation}

The overall optimization problem can then be formulated as the maximization of $R$ within the set of admissible policies. Unfortunately, unlike in the case $U=1$, the problem is non-convex, and therefore the global maximum cannot be easily found. Through a game theoretical formulation, the problem can be addressed as follows: (\emph{i}) we assume that the devices play a game, each trying to increase its own reward; (\emph{ii}) we look for Nash Equilibria in this game, i.e., situations in which no user has an incentive to deviate; (\emph{iii}) in particular, we look for symmetric Nash Equilibria, where all users adopt the same policy and they all have no incentive to do otherwise; (\emph{iv}) it is possible to show that the game has a unique symmetric Nash Equilibrium, and that such equilibrium is a local optimum of the original problem; (\emph{v}) while there is of course no guarantee that the local optimum is also a global optimum, some numerical investigation has shown that the solution found as described above is very good unless the battery capacity of the devices is very small. More details about the above formulation as well as numerical results and discussions can be found
in \cite{MichelusiICC13}.

\section{Medium Access Control for Energy Harvesting Networks}
\label{eh-macprotocols}
In this section, we address medium access control (MAC) protocols for single hop networks in which a \textit{\emph{fusion center}} collects data from energy harvesting devices in its surroundings. We consider the case in which the devices generate data periodically, when timed measurements of a given quantity of interest need to be reported.

We investigate how the performance and design of standard MAC protocols, such as TDMA, framed-ALOHA (FA) and dynamic-FA (DFA) \cite{Schoute}, are influenced by the discontinuous energy availability in the energy harvesting devices in the presence of periodic data generation; see \cite{IannelloTCOM, IannelloCISS, IannelloTAC}. Consider a single hop network with a fusion center surrounded by $M$ energy harvesting devices. The fusion center retrieves data, e.g., measurements of a given phenomenon of interest, from the devices via periodic \textit{inventory rounds} (IRs). Each IR is started by the fusion center by transmitting an initial \emph{query command}, which provides both synchronization and instructions to the devices on how to access the channel. Time is slotted. In every IR, each device has a packet, e.g., a new measurement, to transmit with a given probability, independent of the other devices and previous IRs. Each measurement is the payload of a packet, whose transmission fits within the slot duration. The goal of the fusion center in each IR is to collect as many packets as possible within the constraints imposed by the energy availability at the devices.

Each IR is organized into \textit{frames}, each of which is composed of a number of slots that is selected by the fusion center; see Fig.~\ref{fig:efficiencies_region_fixed_gamma}(a). Depending on the adopted MAC protocol, any device that needs to transmit in a frame either chooses or is assigned a single slot within the frame for transmission. Moreover, after a device has successfully transmitted its packet to the fusion center, it first receives an acknowledgment of negligible duration from the fusion center and then becomes inactive for the remaining of the IR. The fusion center knows neither the number of devices with a new measurement to transmit nor the state of the devices' batteries. Assuming flat fading channels and that transmission is successful if the signal-to-interference ratio is large enough, any slot can be: \textit{empty} when it is not selected by any device; \textit{collided} when it is chosen by more than one device but no device is received successfully; \textit{successful} when a device transmits successfully, while possibly in the presence of other (interfering) devices. Each device has a finite battery, and a fixed amount of energy is consumed at each packet transmission. During the time between two successive IRs, each device harvests a random amount of energy with average, normalized by the transmission energy, defined as $\mu_{H}$. No energy is harvested within an IR, as its duration is assumed to be much smaller than the time needed to harvest sufficient energy for a transmission.

We measure the system performance in terms of the trade-off between the \textit{time efficiency}, which measures the rate of data collection at the fusion center and the \textit{delivery efficiency}, which accounts for the number of packets successfully reported to the fusion center. The time efficiency is a standard measure for the performance of MAC protocols, and is calculated as the ratio between the overall number of packets successfully received by the fusion center and the total number of slots allocated by the MAC protocol. The delivery efficiency is instead the fraction of devices that are able to successfully report their payload to the fusion center within a given IR. This criterion is specifically relevant for energy harvesting networks, since energy harvesting devices may run out of energy before being able to transmit successfully. We observe that, with contention-based MACs such as ALOHA and variations thereof, there is a trade-off between time and delivery efficiencies. Increasing the delivery efficiency requires the fusion center to allocate a larger number of slots in an IR, as this reduces packet collisions and hence the amount of energy wasted in unsuccessful transmissions; however, a larger number of slots per IR decreases the time efficiency.

With the \emph{TDMA protocol}, each device is pre-assigned an exclusive slot in each IR, irrespective of whether it has a packet to deliver or enough energy to transmit. Recall that such information is not available at the fusion center. Any IR is thus composed by one frame with $M$ slots as seen in Fig.~\ref{fig:efficiencies_region_fixed_gamma}(a). Since TDMA is free of communication errors in the considered interference-limited scenario, its delivery efficiency is only limited by the energy availability at the devices and it is thus an upper bound on the delivery efficiency for ALOHA-based MACs. However, TDMA might not be time efficient due to the many empty slots when the probability of having a new measurement and/or the energy harvesting rate are small.

With the \emph{DFA protocol}, the $n$th IR is organized into a set of frames as shown in Fig.~\ref{fig:efficiencies_region_fixed_gamma}(a). The \emph{backlog} for the $k$th frame is the set composed of all sensors that simultaneously satisfy the following three conditions: \textit{i}) have a new measurement to transmit in the $n$th IR; \textit{ii}) have transmitted unsuccessfully, because of collisions, in the previous $k-1$ frames; \textit{iii}) have enough energy left in the battery to transmit in the $k$th frame. All the devices in the backlog set attempt transmission during frame $k$. To make this possible, the fusion center allocates a frame of $L_{k}(n)$ slots, where $L_{k}(n)$ is selected based on an estimate of the backlog size. A typical choice is to make $L_{k}(n)$ proportional to the estimated backlog. We define the proportionality constant as $\rho$. It is known that the optimal $\rho$ for non-energy harvesting devices in terms of time efficiency equals $1$ \cite{Schoute}. However, following the discussion above (see also \cite{IannelloTCOM}), this is no longer the case for energy harvesting enabled networks. Finally, FA is a special case of DFA where only one single frame of size $L_{1}(n)$ is announced and hence the retransmission of collided packets is not allowed.

The trade-off between the delivery efficiency and the time efficiency is shown in Fig.~\ref{fig:efficiencies_region_fixed_gamma}(b) for the considered MAC protocols and for different values of the harvesting rate $\mu_{H}$. For TDMA, the trade-off consists of a single point on the plane, whereas FA and DFA allow for more flexibility via the selection of the parameter $\rho$. Specifically, when $\rho$ is increased more devices can report their measurements to the fusion center, thus increasing the delivery efficiency, but at the cost of lowering the time efficiency.

\begin{figure*}[t]
\centerline{\begin{tabular}{cc}
\subfigure{\includegraphics[width=0.44\linewidth]{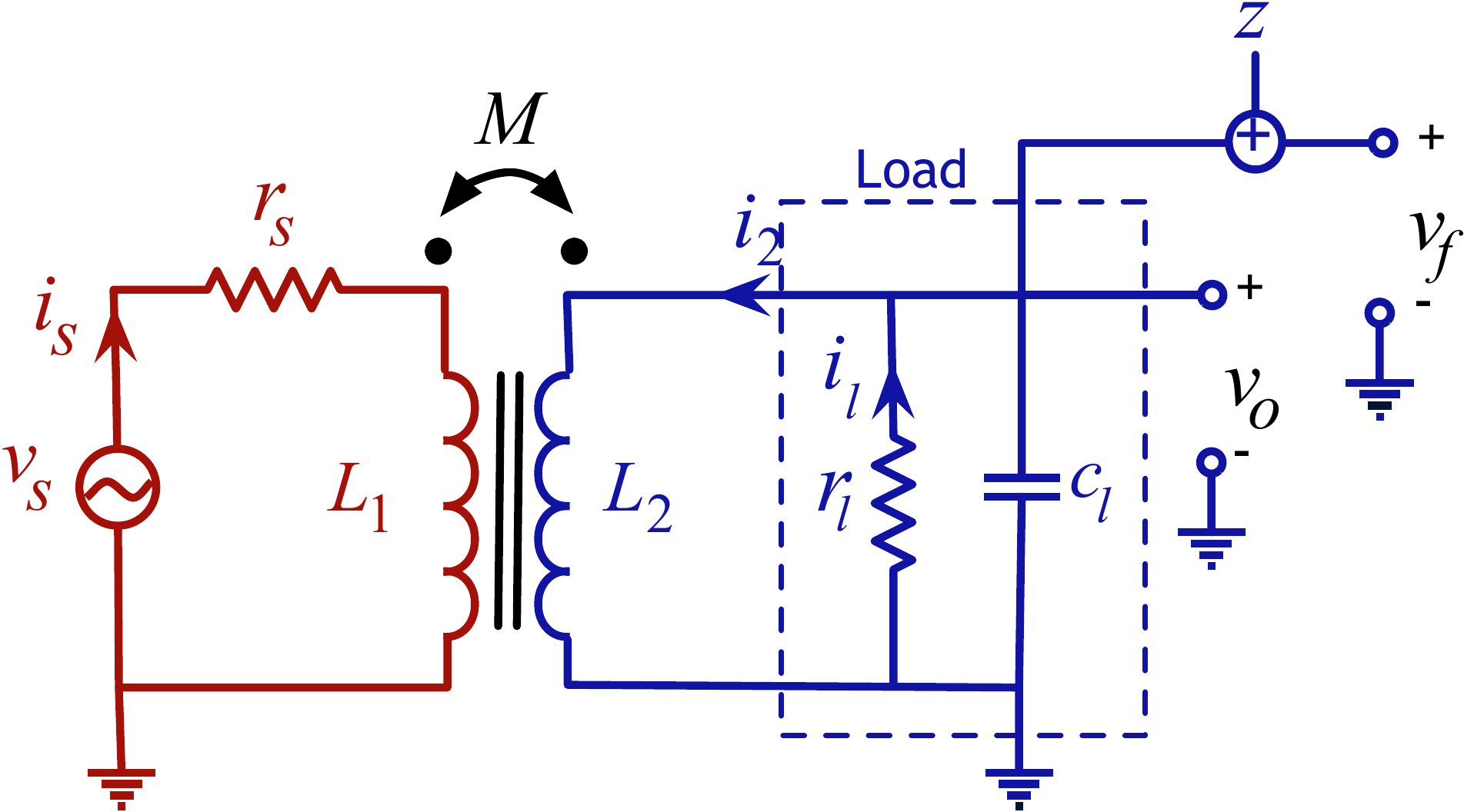}}&
\subfigure{\includegraphics[width=0.35\linewidth]{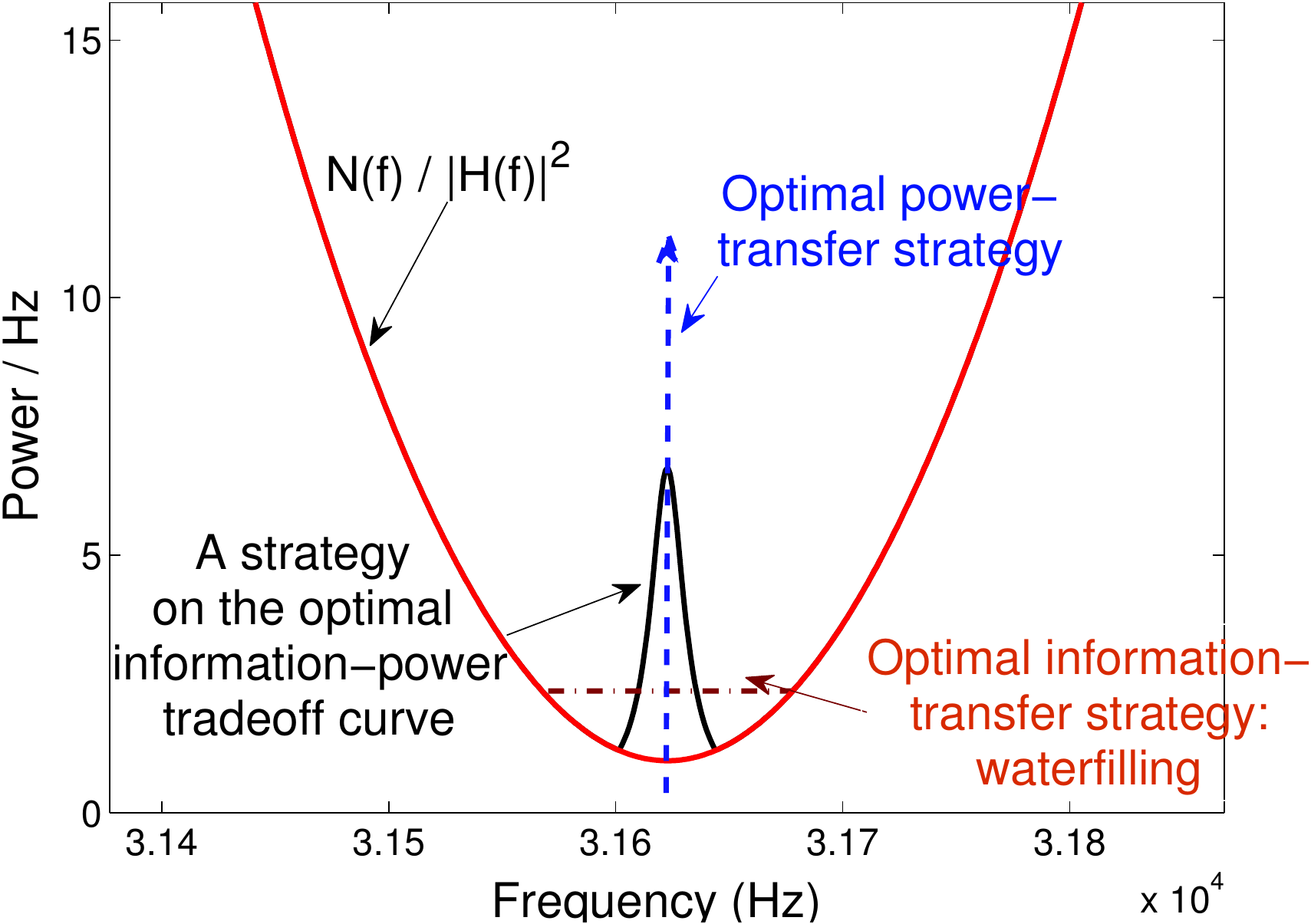}}\\
(a) & (b)
\end{tabular}}
\caption{(a) Coupled-inductor circuit commonly used for wireless power transfer. (b) Frequency-faded channel transfer function.}
\label{fig:coupled}
\vspace*{-0.3cm}
\end{figure*}

\section{Joint Wireless Energy and Information Transfer}
\label{eh-et}
In the preceding sections, we considered wireless nodes that acquire their operational energy from external sources. In this realm of energy self-sustaining networks, a new dimension that can be envisioned is sharing and transferring energy between the nodes just like information. Energy and information transfer between nodes can be made simultaneously or by separate means/technologies. In this section, we review the state of the art in energy and information transfer for wireless networks.

\subsection{An Information-Theoretic View}

The possibility of harvesting energy from the communicated signal itself has been discussed in the circuits literature, especially in the context of biomedical implants~\cite{SarpeshkarBook}. The first explicit information-theoretic formulation of harvesting energy from the communicated signal is given in \cite{Varshney2008}; prior to this work, \cite{gastpar-receive} studied a related problem, where capacity with receiver-side power constraints were considered. Reference \cite{Varshney2008} modeled the problem as one of maximization of mutual information under a minimum requirement of energy delivered to the receiver:
\begin{equation}
C(B)= \max_{f(x)} I(X;Y), \qquad \mbox{s.t.} \quad E[Y^2]\geq B
\end{equation}
where $B$ denotes the energy harvested by the receiver, and $C(B)$ denotes the capacity at this harvesting level. This reference also provided tradeoffs between capacity and average energy delivered for some example cases. The information-energy transfer formulation is also mathematically similar to the classical problem of cost-constrained capacity in \cite[Thm.~6.11]{CsiszarKorner} and \cite[Thm.~3.7.2]{HanBook}. For finite alphabets, by defining the ``cost'' of a particular input symbol as the difference of the energy delivered by the  symbol, and the maximum energy delivered over all symbols, one can reduce the information-energy transfer problem to the cost-constrained capacity problem.

For energy and information transfer using inductively coupled circuits (see Fig.~\ref{fig:coupled}), assuming an additive Gaussian noise, reference \cite{ShannonMeetsTesla} observed that this problem can be viewed as transfer of information and energy over a frequency-faded channel. Fig.~\ref{fig:coupled}(a) shows a coupled-inductor circuit commonly used for wireless power transfer. The same circuit models near-field communication commonly used in RFIDs and medical implants~\cite{SarpeshkarBook}. Fig~\ref{fig:coupled}(b) shows a transfer function of $H(f)$ and white noise $z$ (of power spectral density $N(f)=1$); the curve is the plot of $N(f)/|H(f)|^2$. Reference \cite{ShannonMeetsTesla} derived the tradeoffs between capacity and energy delivery. For this system, the optimal information-transfer strategy is \emph{water-filling} \cite{CoverThomas}; whereas the optimal power-transfer strategy is to send a single sinusoidal tone at the \emph{resonant} frequency. However, the optimal info-power transfer tradeoff is not achieved by a strategy that time-shares between these two. Reference \cite{ShannonMeetsTesla} showed that the optimal strategy outperforms any strategy that time-shares between water-filling and transmission of a single tone at the resonant frequency.

The simultaneous energy and information transfer literature has been extended to MIMO broadcast \cite{revision-reviewer3-1}, fading \cite{revisionL}, MIMO interference \cite{revision-reviewer3-2} channels, and also considering practical circuit implementations \cite{rui-zhang-power-split} keeping in mind the current practical limitations on receivers which may not be able to decode information and harvest energy simultaneously.

\subsection{Coding for Energy and Information Transfer}

In this section, we concentrate on the problem of \emph{code design} for systems with joint information and energy transfer. Classical codes, which are designed with the only aim of maximizing the information rate, are unstructured, i.e., random-like. As a result, they do not allow to control the timing of the energy transfer. This is a critical drawback, since, in most scenarios of interest, arbitrary energy transmission patterns may lead to inefficiencies, such as battery overflows or underflows. In contrast, as reviewed here, constrained, rather than classical unconstrained, codes allow the energy transfer properties of the code to be better adjusted to the receiver's energy utilization requirements \cite{FouladgararXiv}.

We consider a point-to-point link and assume binary transmission, in which ``1'' symbols carry energy while no energy is carried by ``0'' symbols. Barring channel losses, which happen with a given probability, the receiver can harvest the energy carried by the ``1'' symbols and store it in a battery. The receiver's energy utilization is modeled as a binary Markov stochastic process. Due to the finite capacity of the battery, there may be battery overflows and underflows. An overflow event takes place when energy is received but the battery is full; instead, an underflow event occurs when energy is required by the receiver but the battery is empty. Note that the probability of overflow measures the efficiency of energy transfer by accounting for the energy wasted at the receiver. In contrast, the probability of underflow is a measure of the fraction of the time in which the application run at the receiver is in outage due to the lack of energy.

\begin{figure*}[t]
\centerline{\includegraphics[width=0.68\linewidth]{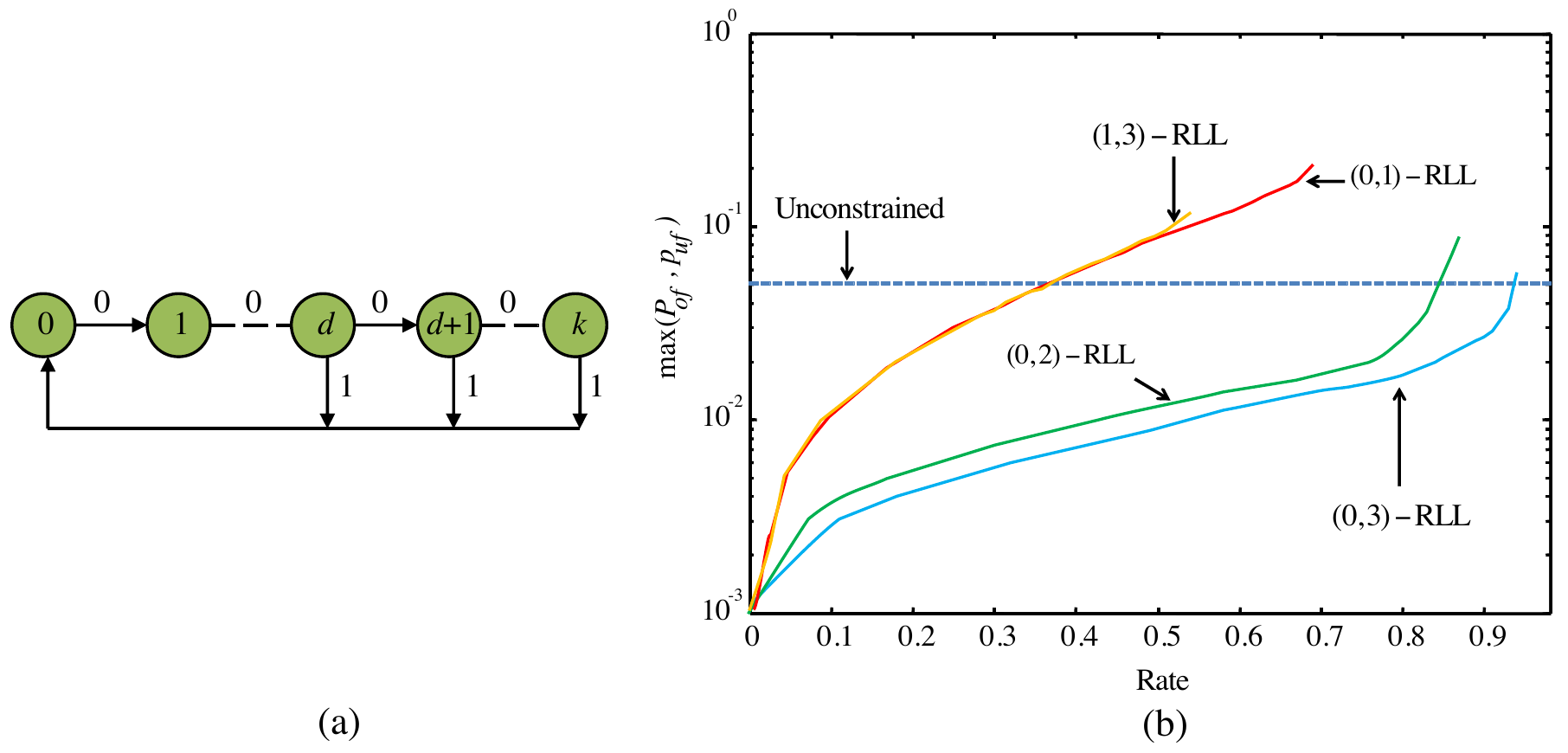}}
\caption{(a) Finite-state machine that defines the constraints that the codewords
of a type-0 $(d,k)$-RLL codes must satisfy. (b) Maximum between
probability of underflow and overflow for unconstrained and type-0
constrained codes versus the information rate.}
\label{fig:fig12}
\vspace*{-0.3cm}
\end{figure*}

Constrained run-length limited (RLL) codes are defined by constraints on the minimum and maximum duration of bursts of ``1'' or ``0'' symbols. Specifically, type-0 $(d,k)$-RLL codes are such that the runs of 0s have length at most $k$, while the runs of 0s between successive 1s have length at least $d$; see Fig.~\ref{fig:fig12}(a). As a result, type-0 $(d,k)$-RLL codes are suitable for overflow-limited regimes in which controlling overflow events is most critical. Type-1 $(d,k)$-RLL codes are similarly defined by substituting ``1'' for ``0'', and are hence appropriate for underflow-limited regimes in which it is necessary to ensure the presence of bursts of energy. Constrained RLL codes have been traditionally studied for applications related to magnetic and optical storage \cite{Marcus}. The first application to the problem of energy transfer has been reported in the context of point-to-point RFID systems in \cite{Barbero}.

We now provide a numerical example to compare the performance of unconstrained and constrained codes. In Fig.~\ref{fig:fig12}(b), we observe the trade-off between the information rate and the performance in terms of energy transfer. The latter is measured by the minimum between the probability of underflow $P_{uf}$ and the probability of overflow $P_{of}$. In the example, the receiver wishes to use energy periodically once every two time slots. A full description of the simulation set-up can be found in \cite{FouladgararXiv}. The figure shows that, when the desired rate is small, it is sufficient to use a type-0 RLL code with a small $k$, since, with this choice, the resulting pattern of 0s and 1s in the codewords matches well the periodic requests of energy by the energy harvesting receiver. The resulting improvement in terms of energy transfer efficiency is significant. As the rate grows larger, one needs to increase the value of $k$, while keeping $d$ as small as possible in order to increase the number of available codewords and hence the rate \cite{Marcus}.

\begin{figure*}[t]
\centerline{\begin{tabular}{cc}
\subfigure{\includegraphics[width=0.45\linewidth]{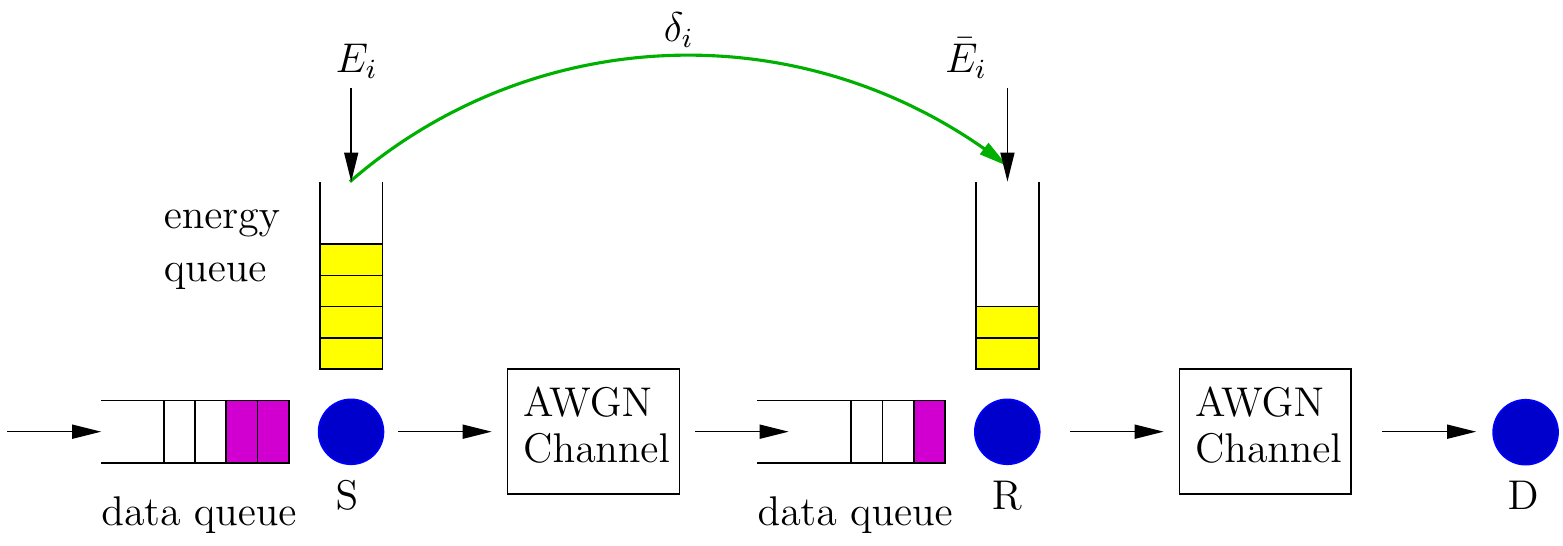}}&
\subfigure{\includegraphics[width=0.35\linewidth]{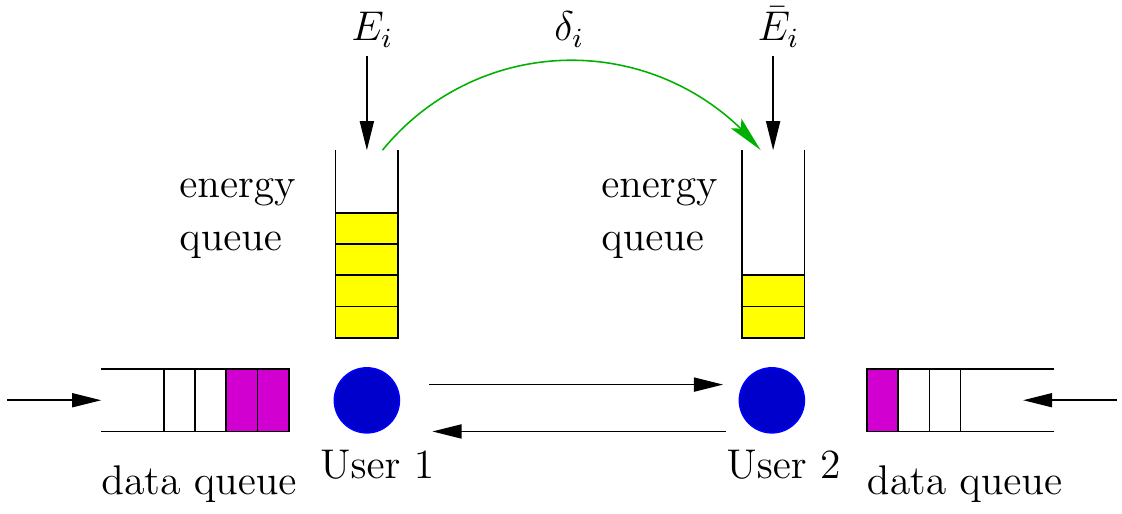}}\\
(a) & (b)
\end{tabular}}
\caption{Energy cooperation via wireless energy transfer. (a) Relay channel. (b) Two-way channel.}
\label{ener-coop-fig1}
\vspace*{-0.3cm}
\end{figure*}

\subsection{Energy Cooperation}

Consider a three-node relay network as shown in Fig.~\ref{ener-coop-fig1}(a). Here, both the source node and the relay node harvest energy from nature in the amounts $E_i$ and $\bar{E}_i$ in slot $i$. In addition, there is a wireless energy transfer unit that transfers $\delta_i$ amount of energy from the source's battery into the relays battery in slot $i$. When the source node sends $\delta_i$ amount of energy, the relay node receives $\alpha \delta_i$ amount of energy into its battery, where $\alpha<1$ accounts for the inefficiency of wireless energy transfer. The goal of the network is to deliver the data packets of the source node to the destination node, and in the process, maximize the end-to-end throughput of the system, from the source node to the destination node.

The problem of power control to maximize the end-to-end data throughput of this relay system without any energy transfer was studied in \cite{gunduz-camsap, Zhang_Relay}. They showed that the optimum scheme should work in the following way: The source node should transmit as many data packets as possible to the relay node by using its harvested energy. This will cause a certain packet arrival profile at the relay node. The relay node then should transmit as many of these data packets as possible to the destination node by using its harvested energy. One can note now that some of the packets that were delivered by the source node to the relay node may not be forwarded to the destination node if the relay node does not have a sufficient amount of harvested energy. Therefore, some of the data packets delivered to the relay node will have no utility for the overall system (i.e., will not contribute towards the end-to-end throughput) since they are not delivered to the destination node. Note that the source node had spent some of its energy to deliver these eventually undelivered data packets to the relay node.

One could think that the source node might as well not transmit those data packets to the relay node and keep its remaining energy. However, a better solution would be obtained if the source node sent some of its remaining energy to the relay node via wireless energy transfer and used the rest of the remaining energy to send as many data packets as possible to the relay node. The source node will determine the amount of energy to be transferred to the relay node such that there will be no remaining energy and no remaining data packets at the relay node at the end of the transmission session. This is the rationale behind the concept of \emph{energy cooperation} introduced in \cite{gurakan12isit}; see also \cite{gurakan13tcom}. In this scheme, we observe two types of cooperation: the relay node cooperates at the {\it signal level} by forwarding the source's data packets to the destination \cite{coop_div}, and the source node cooperates at the {\it energy level} by transferring some of its harvested energy to the relay node. These two cooperation schemes combined yield the optimum scheme for the overall system. Note that, even though there was energy transfer inefficiency in the system represented by $\alpha<1$, so long as the source node had sufficiently more energy than the relay node, it is worth transferring some of the source energy to the relay.

\begin{figure*}[t]
\centerline{\begin{tabular}{cc}
\subfigure{\includegraphics[width=0.22\linewidth]{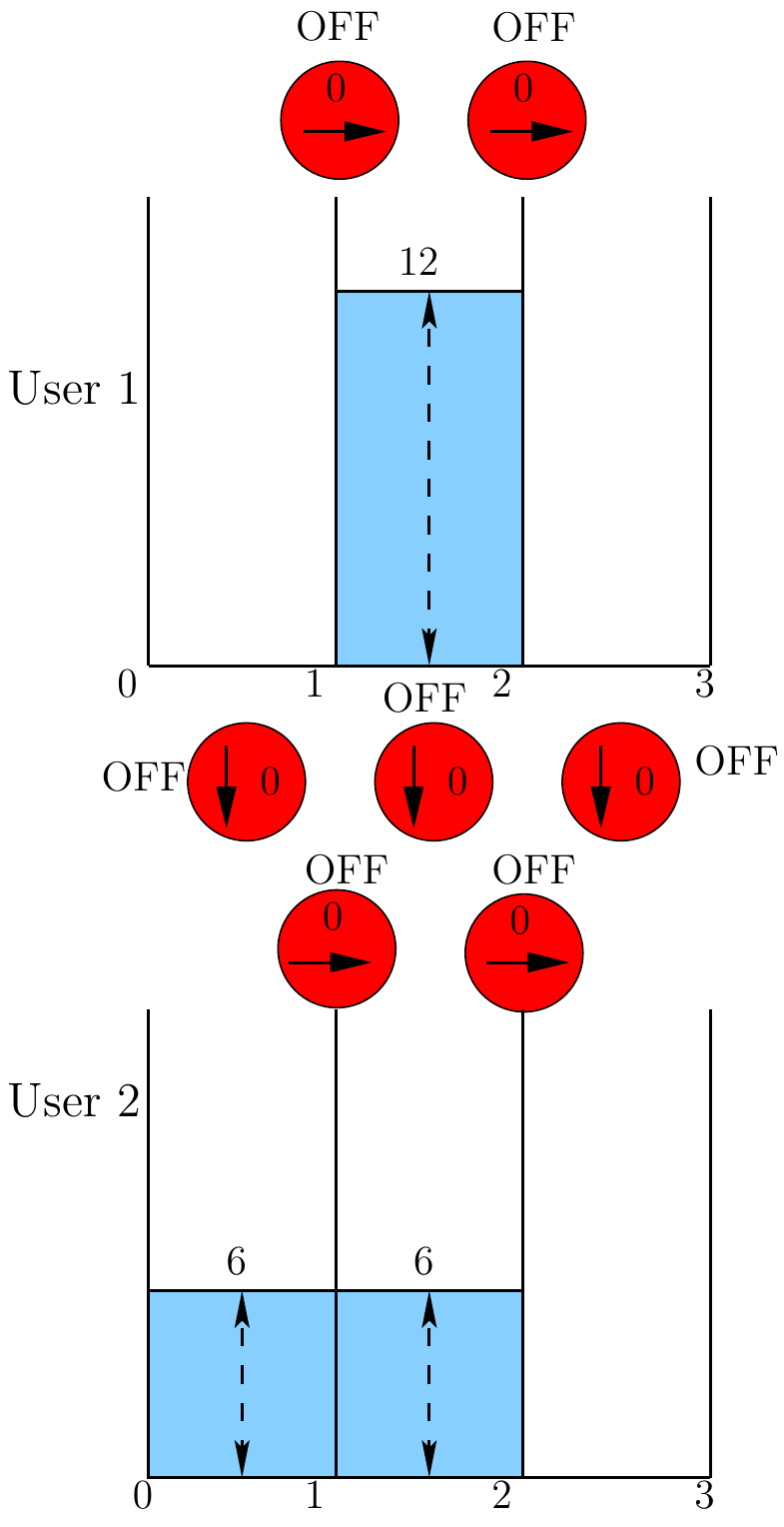}}& \qquad
\subfigure{\includegraphics[width=0.22\linewidth]{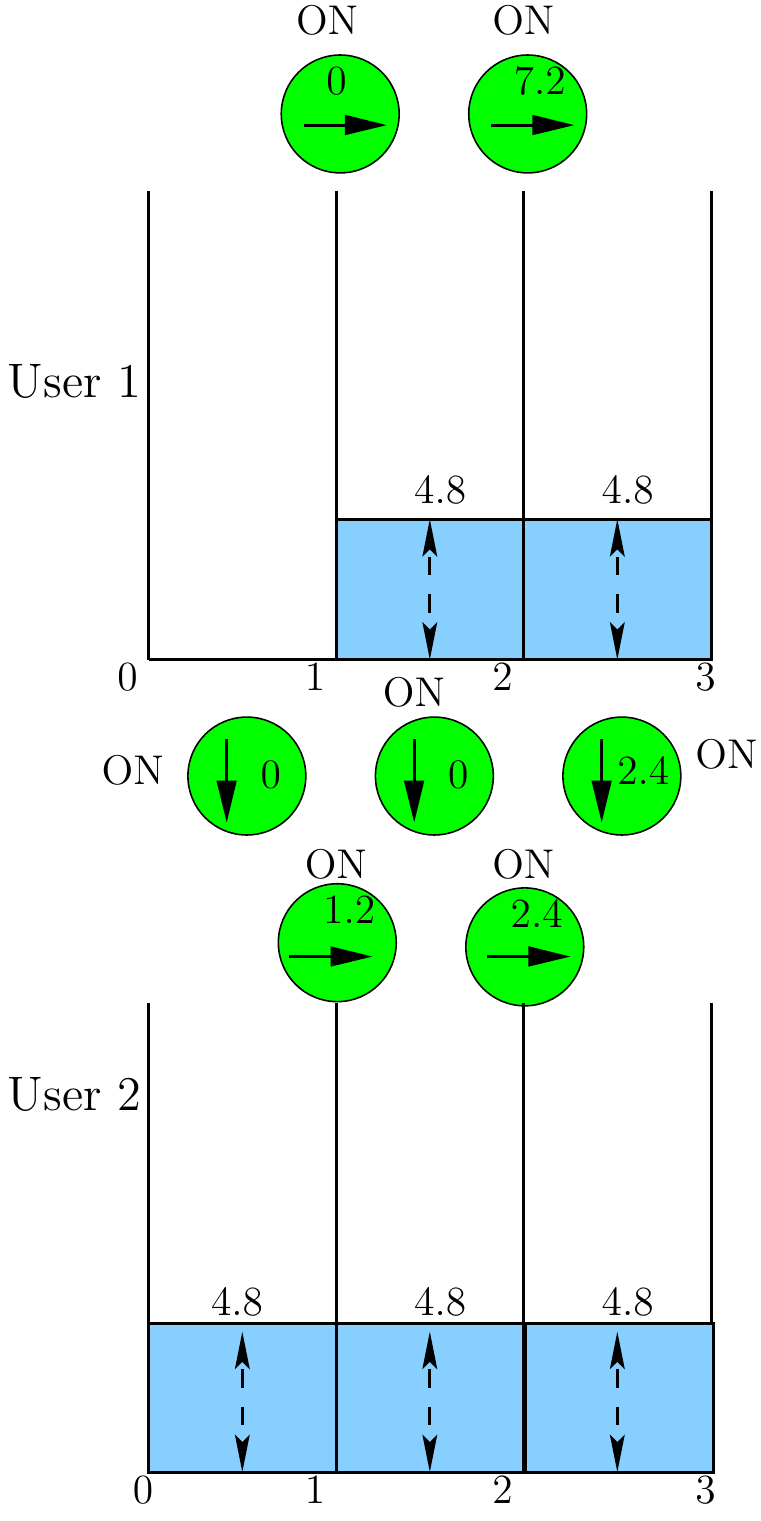}}\\
(a) & \qquad (b)
\end{tabular}}
\caption{Two-dimensional directional water-filling. (a) Initial energy allocation. (b) Final energy allocation.}
\label{ener-coop-fig2}
\vspace*{-0.3cm}
\end{figure*}

The concept of energy cooperation in two-way and multiple access channels have been studied in \cite{gurakan12asilomar, gurakan13icc}, where users transmit information over two-way and multiple access channels, and there is one-way energy transfer from one user to the other, as motivated by practical scenarios such as RFID networks, where there is two-way information exchange but one-way energy transfer from the reader to the RFID node. References \cite{gurakan12asilomar, gurakan13icc} develop \emph{two-dimensional directional water-filling}, where energy is distributed directionally due to energy causality and flows only from the past to the future and from the energy transferring user to the energy harvesting user. This creates two dimensions for energy flow, in time and over users; and also directionality due to energy causality. Fig.~\ref{ener-coop-fig2} shows an example run of two-dimensional directional water-filling algorithm; (a) shows the initial energy allocation and (b) shows the final energy allocation. More recent works \cite{tutuncuoglu2013ita, tutuncuoglu2013itw} generalize this approach to consider the case with two-way energy cooperation in two-way and multiple access channels, and develop a separation-based approach that optimizes wireless energy transfer and temporal power allocation separately to yield a global optimum solution for the overall problem. As a final remark, we note that in \emph{energy cooperation} described in this section, energy and information are sent by different signals over orthogonal channels. This is reminiscent of the \emph{power splitting} approach implemented at the receivers, see for example \cite{rui-zhang-power-split}, however, we implement power splitting at the transmitter as opposed to the receiver.

\subsection{Interactive Exchange of Energy and Information}

\begin{figure*}[t]
\centerline{\includegraphics[width=0.65\linewidth]{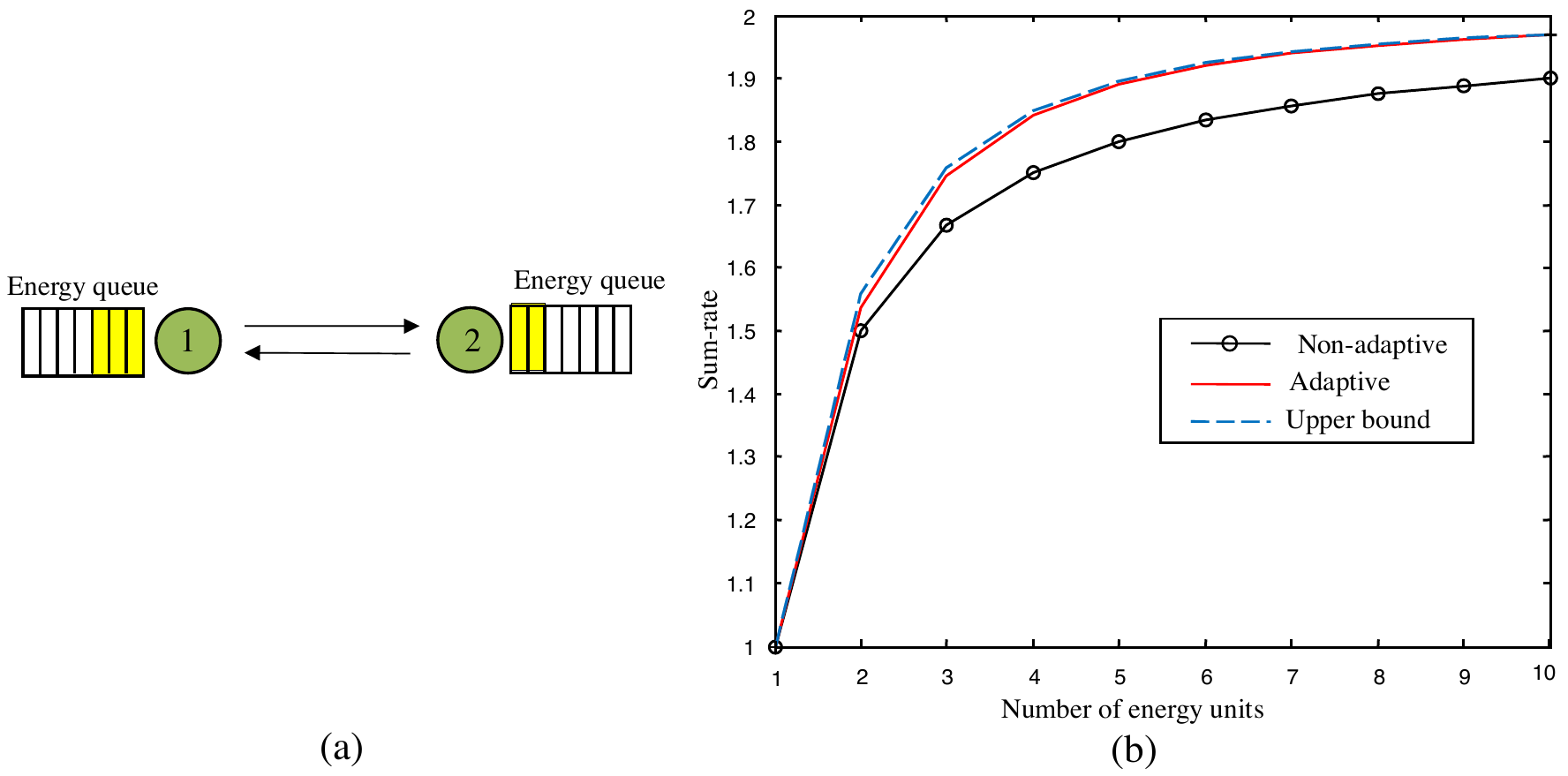}}
\caption{(a) Two-way system with energy reuse. (b) Achievable sum-rates with
non-adaptive and adaptive codebooks and upper bound.}
\label{fig:num_results}
\vspace*{-0.3cm}
\end{figure*}

In this section, we consider a multi-hop topology in which the harvested energy can be reused for communications. Multi-hop networks with distinct source and destinations are considered under this assumption in \cite{FouladgarWCL}. Here, instead, we study a two-way communication system, in which two nodes interact for the exchange of information and can harvest the received energy; see Fig.~\ref{fig:num_results}(a). To enable analysis and insights, we assume that the two parties involved have a common clock and that, at each time, a node can either send a ``1'' symbol, which carries one unit of energy, or a ``0'' symbol, which does not carry any energy. Each node communicates in a full-duplex manner, that is, at a given instant, it can simultaneously send and receive an energy unit. The channel in one direction is orthogonal to the channel in the other direction, and hence the full-duplex channel is an ideal composition of two independent unidirectional channels. We consider here the case in which the two nodes start with a given number of energy units in their batteries, which can neither be lost or replenished from outside, and the binary channel in either direction is noiseless. Extensions
can be found in \cite{PopovskiTCOM}.

To see that even this simple scenario offers relevant research challenges, we observe the following. If there were no limitation on the number of energy units, the nodes could communicate 1 bit per channel use in either direction, given that the channels are ideal. However, if there is, say, one energy unit available in the system, only the node that currently possesses the energy unit can transmit a ``1'', whereas the other node is forced to transmit a ``0''. Therefore, the design of the communication strategy at the nodes should aim not only at transferring the most information to the counterpart, but also to facilitate energy transfer to enable communication in the reverse direction. In \cite{PopovskiTCOM}, a coding strategy is proposed for this set-up that employs codebook multiplexing. The basic idea is that each node utilizes a different codebook for each energy state. An energy state is defined by the current distribution of the energy units between the two nodes. Note that, under the given assumptions, both nodes are aware of the current energy state. Whenever a given energy state takes place, each node sends the next bit from the corresponding codebook. The key point is that, when the number of available energy units at the node is large, the node should use a codebook with a larger fraction of ``1'' symbols in order to facilitate energy transfer; instead, when the available energy is scarce, a codebook with a larger fraction of ``0'' symbols should be used.

Fig.~\ref{fig:num_results}(b) compares the achievable sum-rate obtained with the mentioned scheme to an upper bound derived in \cite{PopovskiTCOM} versus the total number of energy units. Specifically, for the achievable
sum-rate, we consider both a conventional codebook design, in which all the codebooks have the same fraction of 0s and 1s irrespective of the energy state (labeled as ``non-adaptive'' in the figure), and one in which the probabilities are optimized (labeled as ``adaptive'' in the figure). It can be seen that using conventional codebooks, which only aim at maximizing information flow on a single link, leads to substantial performance loss. Instead, the proposed strategy with optimized probabilities, which account also for the need to manage the energy flow in the two-way communication system, performs close to the upper bound. The latter is indeed achieved when the number of energy units is large enough.

\section{Energy Harvesting Wireless Sensor Networks}
\label{ehsn}
Wireless sensor networks differ from the wireless systems considered thus far in that the devices need not only to cater to the requirements of data transmission but also to those of source acquisition. Specifically, each sensor runs a source acquisition that involves sensing, sampling and compression, and these operations often entail an energy cost that is comparable with that of radio transmission; see, e.g., \cite{HE06, LU03, revision N, revisionO}. Therefore, a proper allocation of the limited energy resources to source acquisition and transmission is necessary. As a result, the techniques studied above that only adapt to the temporal variations of the energy harvesting process and of the transmission channel must be revised in order to account for the time-varying properties of the source acquisition systems, e.g., for the quality of the measurements taken by the sensor.

\begin{figure*}[t]
\centerline{\includegraphics[width=0.9\linewidth]{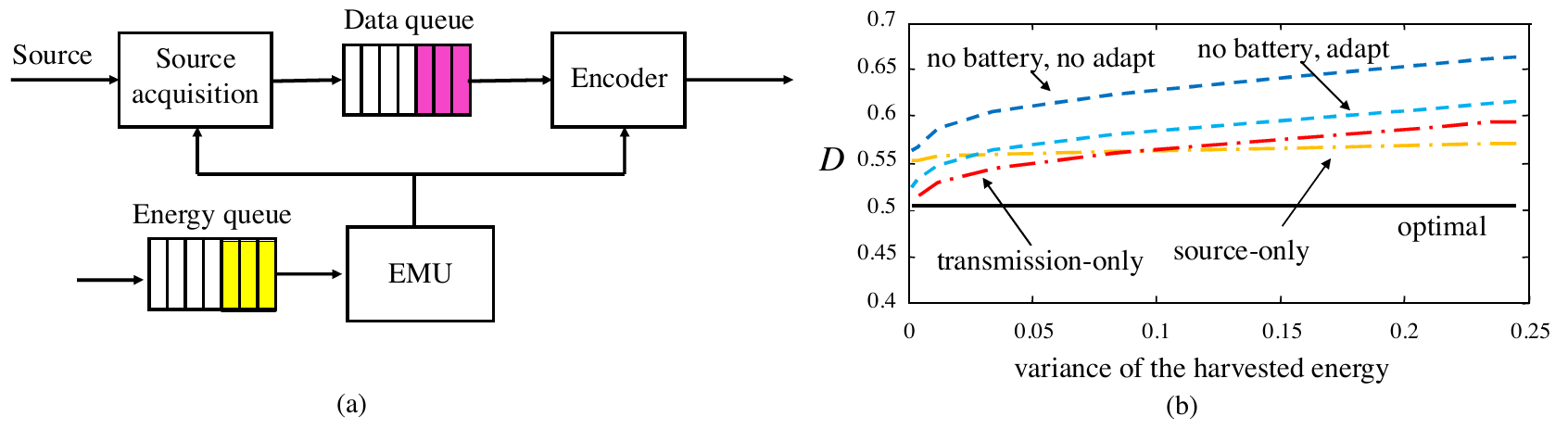}}
\caption{(a) System model for an energy harvesting wireless sensor link; (b) Average distortion
$D$ of the source reconstruction at the receiver versus the variance of the harvesting process.}
\label{fig:castiglione1}
\vspace*{-0.3cm}
\end{figure*}

In order to concentrate on the main aspects of the problem, we focus on a system, studied in \cite{CastiglioneTCOM}, in which a single sensor communicates with a single receiver. Time is slotted. The energy harvested in each time slot is assumed to follow an ergodic stationary process. As shown in Fig.~\ref{fig:castiglione1}(a), the sensor is equipped with a battery (energy queue) in which the harvested energy is stored. The battery is assumed here to be of infinite size for simplicity of analysis. We observe that the case with multiple sensors is studied in \cite{CastiglioneTCOM}, while multi-hop sensor networks are addressed in \cite{Tapparello}. Moreover, a scenario with delay constraints is investigated in \cite{Orhanetal}, where the optimal offline resource allocation policy is derived.

In each time slot, the sensor acquires a time sequence of the phenomenon of interest. This is characterized by a \emph{measurement SNR,} which evolves across the time slots as an ergodic stationary process. The measurements are compressed and stored into a data queue. The quality of compression is defined by a distortion measure, such as the mean squared error. Following \cite{HE06, LU03}, we assume that the bit rate produced by this source acquisition step depends on the energy allocated for source acquisition, on the desired distortion and on the measurement SNR. In every time slot, the sensor also transmits a number of bits from the data queue to the receiver over a fading channel with a given instantaneous \emph{channel SNR}. The channel SNR evolves across the time slots according to a ergodic stationary process. The number of bits that are successfully transmitted depends on the energy allocated to data transmission and on the channel SNR as per, e.g., Shannon's channel capacity.

The energy management problem of interest is the following. Based on the statistics of the energy harvesting process, and based on the current states of the measurement SNR, channel SNR and data queue, the energy management unit (EMU) decides the compression distortion and how much energy to allocate to source acquisition and data transmission (Encoder). The performance criterion considered here is the stability of the data queue, which separates the data acquisition unit from the data transmission block, under a constraint on the average distortion of the measurements recovered at the receiver.

For a given desired average distortion level, we wish to identify a class of policies that is able to stabilize the data queue while satisfying the distortion constraint whenever possible. Note that this definition of optimality generalizes the classical notion of ``throughput optimal'' policies, where only the stability constraint is imposed. As shown in \cite{CastiglioneTCOM}, a class of optimal strategies exists that performs separate resource allocation optimizations for the source acquisition and data transmission. More precisely, without loss of optimality, the battery can be divided into two subcomponents, one used for source acquisition and one for data transmission: the energy harvested in each slot is split according to a fixed factor between the two batteries. Moreover, the energy allocated to source acquisition from the corresponding battery in a given time slot only depends on the measurement SNR, and not on the channel SNR; and, similarly, the energy used for data transmission depends only on the channel SNR.

Fig.~\ref{fig:castiglione1}(b) shows the average distortion of the source reconstruction at the receiver versus the variance of the harvesting process; see \cite{CastiglioneTCOM} for a full description of the set-up. The performance of the optimal policy discussed above is compared to a number of suboptimal policies that either use the energy in a greedy fashion or do not perform adaptation to the current SNR states. Specifically, we consider strategies that use immediately all the harvested energy with and without optimal energy allocation between source acquisition and transmission depending on the SNR states (``no battery'' and ``no battery, adapt'', respectively); and strategies that use the battery to adapt the operation only of source acquisition (``source-only'') or of data transmission (``transmission-only''). It can be seen that an increased uncertainty about the harvested energy, i.e., a larger variance, degrades the quality of the reconstruction at the receiver. Moreover, adapting the energy usage to the current measurement and channel SNRs, especially when leveraging the possibility to store energy in the battery, leads to significant performance gains.

\section{Large-Scale Wireless Networks with Energy Harvesting}
\label{lsehn}
The preceding sections focus on algorithms and protocols for point-to-point links or small-scale systems powered by energy harvesting. From the perspective of network designers and operators, it is interesting to study the effects of energy harvesting on large-scale networks based on models that capture network architectures, node distributions and realistic channel/interference models, which is the theme of this section. In particular, we model mobile ad hoc networks and cellular networks powered by energy harvesting and relate their performance to the characteristics of energy arrival processes or energy fields. Typical renewable energy fields  such as wind and solar power  exhibit both temporal and spatial variations and are commonly modeled as random fields \cite{Cressie1993}. Preceding sections address the temporal variation and how to counteract it by adaptive transmission. On the other hand, the spatial variation of renewables can significant degrade network coverage and hence has to be accounted for in network modeling and  design. To this end, a tractable  model for renewable-energy fields is discussed in the sequel and applied to investigate coverage of renewables powered  networks.

\subsection{Mobile Ad Hoc Networks with Energy Harvesting}\label{Section:MANET}

The spatial throughput of a MANET with energy harvesting is analyzed in \cite{Huang:WirelessAdHocNetworkEnergyHarvesting}.  The transmitters are modeled as a homogeneous Poisson point process (PPP) with density $\lambda_0$ in the horizontal plane. Each communicates with an affiliated receiver located at a unit distance. Given a fixed  encoding rate $\log(1+ \theta)$, reliable decoding of a data packet at a receiver requires the receive SINR to exceed a threshold $\theta$ except for a small outage probability, called an  \emph{outage constraint}. The transmission powers are assumed to be identical and equal to $P$.  Nodes in the MANET are small devices like sensors and wearable computing devices. Given their deployment environment, the energy arrival processes at different transmitters are assumed to be independent i.i.d.~random sequences with mean $\lambda_e$ called the \emph{energy-arrival rate}. For simplicity, the batteries  of harvesters are assumed to have  infinite capacity.

In the steady state, depending on the energy availability, each transmitter is turned on or off with probability $\rho$ and $(1-\rho)$, respectively. The probability is shown in \cite{Huang:WirelessAdHocNetworkEnergyHarvesting} to have a simple form: $\rho = \min\left(1, \lambda_e/P\right)$.  This allows the network throughput to be written as
\begin{equation}
R = \lambda_0 \rho \log(1+\theta)
\end{equation}
Note that $\lambda_0\rho$ is the active-transmitter density. The expression of $\rho$ suggests that the density can be controlled by varying $P$. On one hand, large density and power can cause strong interference and as a result violate the outage constraint. On the other hand, too sparse transmitters reduce  network throughput and too low power leads to incorrect decoding. Thus, transmission power should be optimized under the outage constraint and the criterion of maximum throughput. As shown in \cite{Huang:WirelessAdHocNetworkEnergyHarvesting}, for relatively sparse networks with a sufficiently high energy arrival rate, the optimal power is one that allows all transmitters to be active with probability one. Otherwise, the probability should be  smaller than one as derived.

\subsection{Cellular Networks with Energy Harvesting} \label{Section:Cellular}
\begin{figure*}[t]
\centerline{\begin{tabular}{cc}
\subfigure{\includegraphics[width=0.4\linewidth]{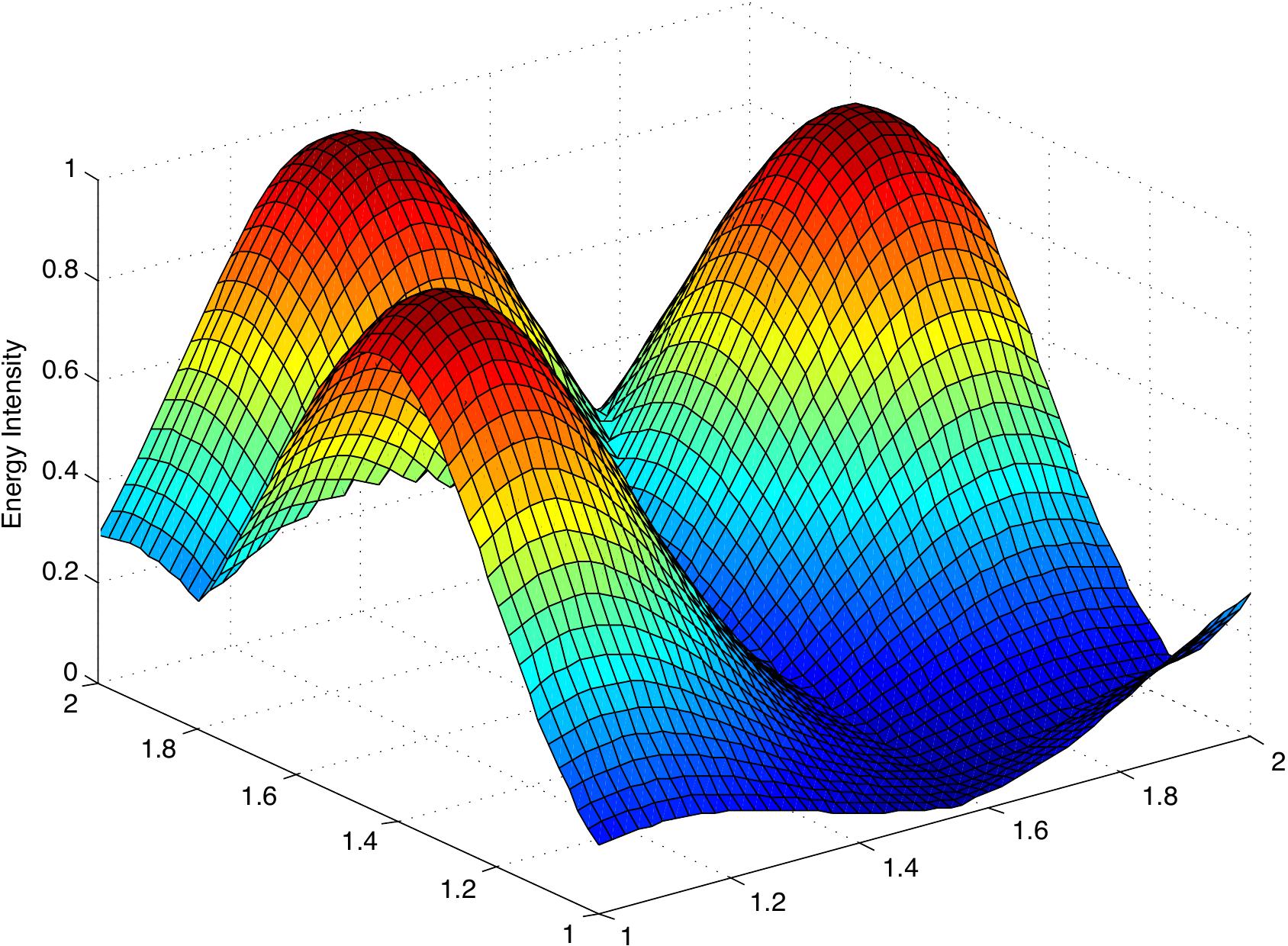}}&
\subfigure{\includegraphics[width=0.4\linewidth]{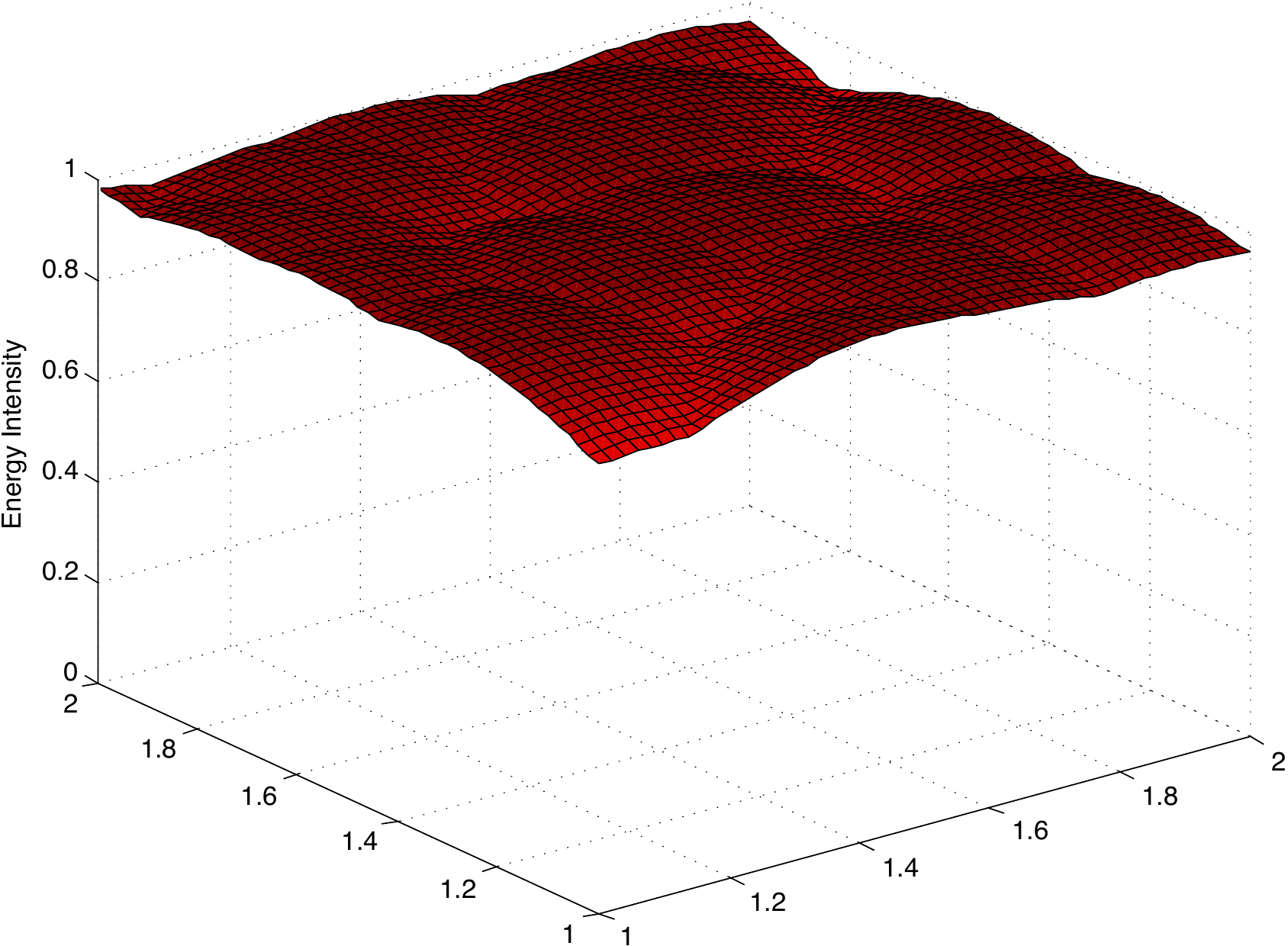}}\\
(a) & (b)
\end{tabular}}
\caption{The energy field. (a) $\lambda_e = 1$ and $\nu = 0.1$. (b) $\lambda_e = 10$ and $\nu = 1$.}
\label{fig:energyfield}
\vspace*{-0.3cm}
\end{figure*}

Reference \cite{HuangLi:RenewablesCellular:2014} addresses the effect of the spatial variation of the (renewable) energy field, e.g., solar or wind power,  on the coverage of a cellular  network. To this end, a tractable energy field model is proposed in  \cite{HuangLi:RenewablesCellular:2014} where the energy intensity at a particular location is given by spatial combining of Poisson distributed ``energy centers" with fixed maximum intensities, which is known as a \emph{Boolean random function}. The combining factors are determined by an exponential-decay function of squared distance. Such a function is commonly used in spatial interpolation for atmospherical mapping \cite{Barnes:NumericalWeatherMap:1964} and solar-field estimation \cite{Sen:SolarEnergy:2004}. The advantage of the energy-field model is that its distribution is controlled by only two parameters,  namely the energy-center density denoted as  $\lambda_e$ and the exponential rate of the weight function denoted as $\nu$. It can be  observed from the illustrations in Fig.~\ref{fig:energyfield}  that the energy field is almost flat for large $\lambda_e$ and $\nu$ or otherwise highly random. Next, the  downlink network is modeled using the traditional hexagonal-cell model but with renewables powered BS's. The network is assumed to operate in  the noise-limited regime that is most interesting for renewables powered network since network performance in the regime  is sensitive to transmission power and hence harvested energy. In addition, an outage constraint is applied on downlink transmissions. We consider two scenarios of harvester deployments. First, each  BS is powered by a single \emph{on-site} harvester.  Based on the energy field model, the outage probability can be related  to the coverage probability of a Boolean model comprising random disks. The result shows that the outage probability decreases exponentially with the product $\lambda_e\nu$.

%\begin{figure*}[t]
%\centerline{
%\subfigure{(a)}{\includegraphics[width=0.40\linewidth]{field_random}}
%\subfigure{(b)}{\includegraphics[width=0.40\linewidth]{field_flat}}}
%\caption{The energy field. (a) $\lambda_e = 1$ and $\nu = 0.1$. (b) $\lambda_e = 10$ and $\nu = 1$.}
%\label{fig:energyfield}
%\vspace*{-0.3cm}
%\end{figure*}

Next, consider the scenario where energy harvested by dense harvesters  is  collected by aggregators that  distribute power to nearby BS's. The aggregators effectively perform spatial averaging of the energy field and consequently stabilize the transmission power of BS's. It is shown in \cite{HuangLi:RenewablesCellular:2014} that as the number of energy harvesters connected to a single aggregator increases, the transmission power converges to a constant by  virtue of the law of large numbers. This renders  the outage probability to be either one or zero. Nevertheless, since harvesters cannot afford high-voltage power transmission, total energy-transmission  loss can become significant as the number of harvesters for energy aggregation increases. Such loss can be regulated by increasing the  voltage at harvesters following a derived scaling law.

\subsection{Other Types of Energy Harvesting Networks}

The  interesting idea of opportunistic energy harvesting is explored in \cite{LeeZhangHuang:OppEnergyHarvestingCognitiveRadio:2013} for cognitive radio networks where passive secondary nodes not only opportunistically access the spectrum of primary nodes but also harvest energy from radiation by the latter whenever it is possible. Coexisting networks are modeled as overlaid spatial point processes. Based on this model, the transmission capacity of the secondary network adopting the strategy of opportunistic energy harvesting is characterized and optimized over the node density and transmission power.

A closely related piece of work is presented in \cite{HuangLauArXiv:EnablingWPTinCellularNetworks:2013} for cellular networks where mobiles are wireless recharged by dedicated power stations via  microwave power transfer (MPT).  Also adopting the approach of using stochastic geometry for modeling and analysis, the requirements on the MPT network deployment are analyzed for different MPT technologies.

Last, heterogeneous cellular networks with energy harvesting is modeled and studied in \cite{DhillonAndrews:HetNetEnergyHarvesting} where multi-tier renewables-powered base stations are modeled as independent PPP's. Considering small base stations deployed in an urban area, energy arrival processes for BS's are suitably assumed to be independent unlike the energy field discussed earlier. This allows the on/off states of BS's to be modeled as independent Bernoulli random variables and then their impact on the network coverage performance quantified mathematically using stochastic geometry.

\section{Energy Consumption Models}
\label{consumption}
Because energy-harvesting systems typically operate at short distances (a few meters or less), the energy consumed in transmitter/receiver circuitry can be comparable, or can even dominate, the energy consumed in transmissions. Thus, it is important to understand energy harvesting communications in the context of \textit{total} (transmit + circuit) energy minimization. The difficulty in developing a comprehensive theory of total energy minimization lies in obtaining models for energy consumed in circuitry that are simple enough for analysis, and yet accurate enough to yield relevant estimates of energy consumption. How can we abstract the energy consumed by various possible circuit algorithms, implementation architectures, and implementation technologies? As a first step, the authors in \cite{Glue2008, goldsmithbahai}, model the transmitter and receivers as black-boxes that consume a fixed amount of energy per unit time powered ``on''. Although the formulations in \cite{Glue2008, goldsmithbahai} are considerably different, their conclusions are the same: since keeping systems powered ``on'' consumes circuit energy, transmissions should be ``bursty,'' i.e., both receiver and transmitter are turned ``off'' for some time in order to reduce circuit energy. However, the energy required in transmission increases exponentially with burstiness (because capacity scales logarithmically in power), so the transmission cannot be too bursty. That the optimal burstiness is non-zero is surprising: traditional transmit power analysis \cite{verducapacitycost} for a non-fading channel predicts that the transmission rate should be made as small as possible, and the signals least bursty, for minimum energy consumption

\begin{figure*}[t]
\centerline{\begin{tabular}{cc}
\subfigure{\includegraphics[width=0.35\linewidth]{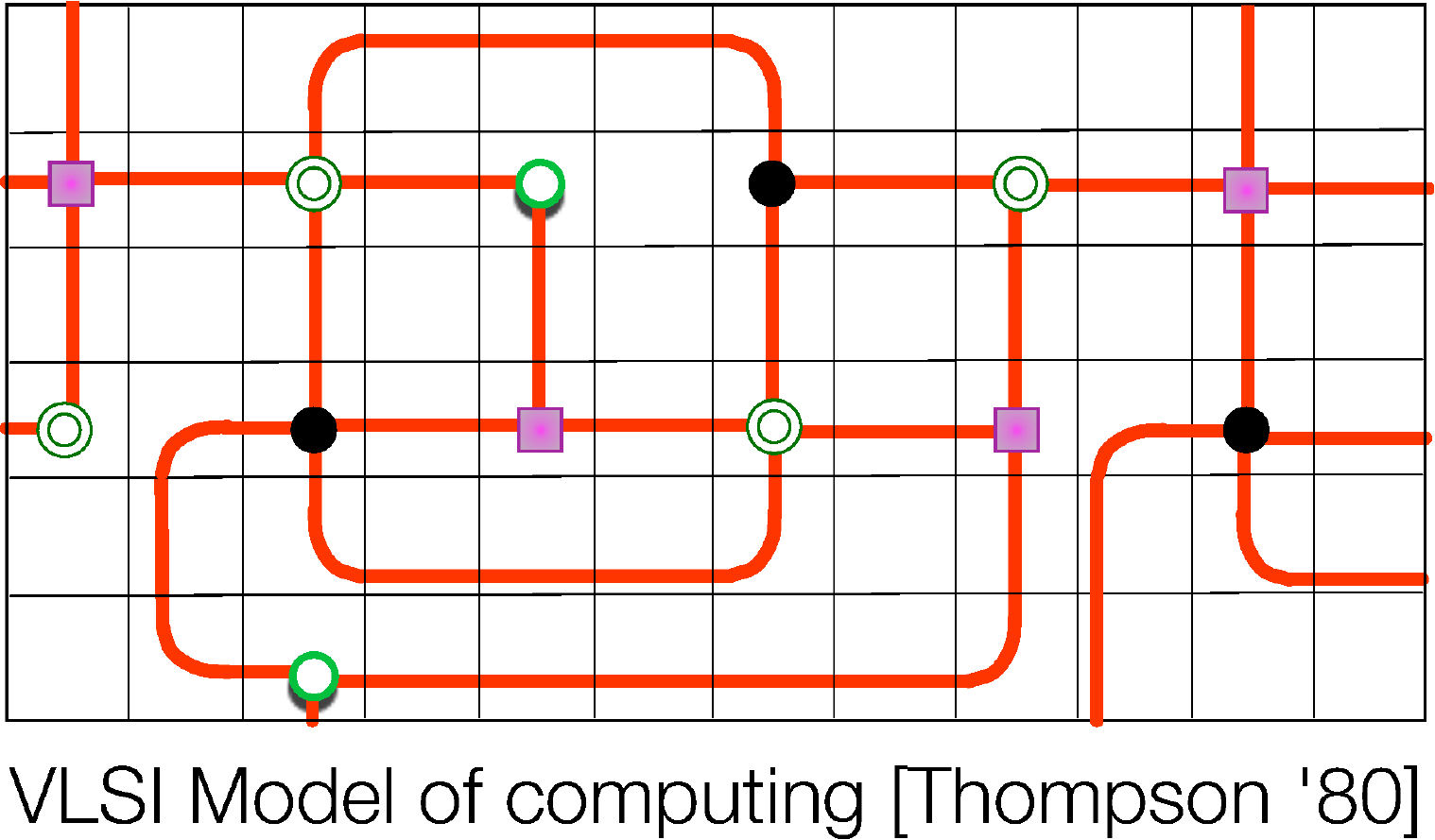}}&
\subfigure{\includegraphics[width=0.45\linewidth]{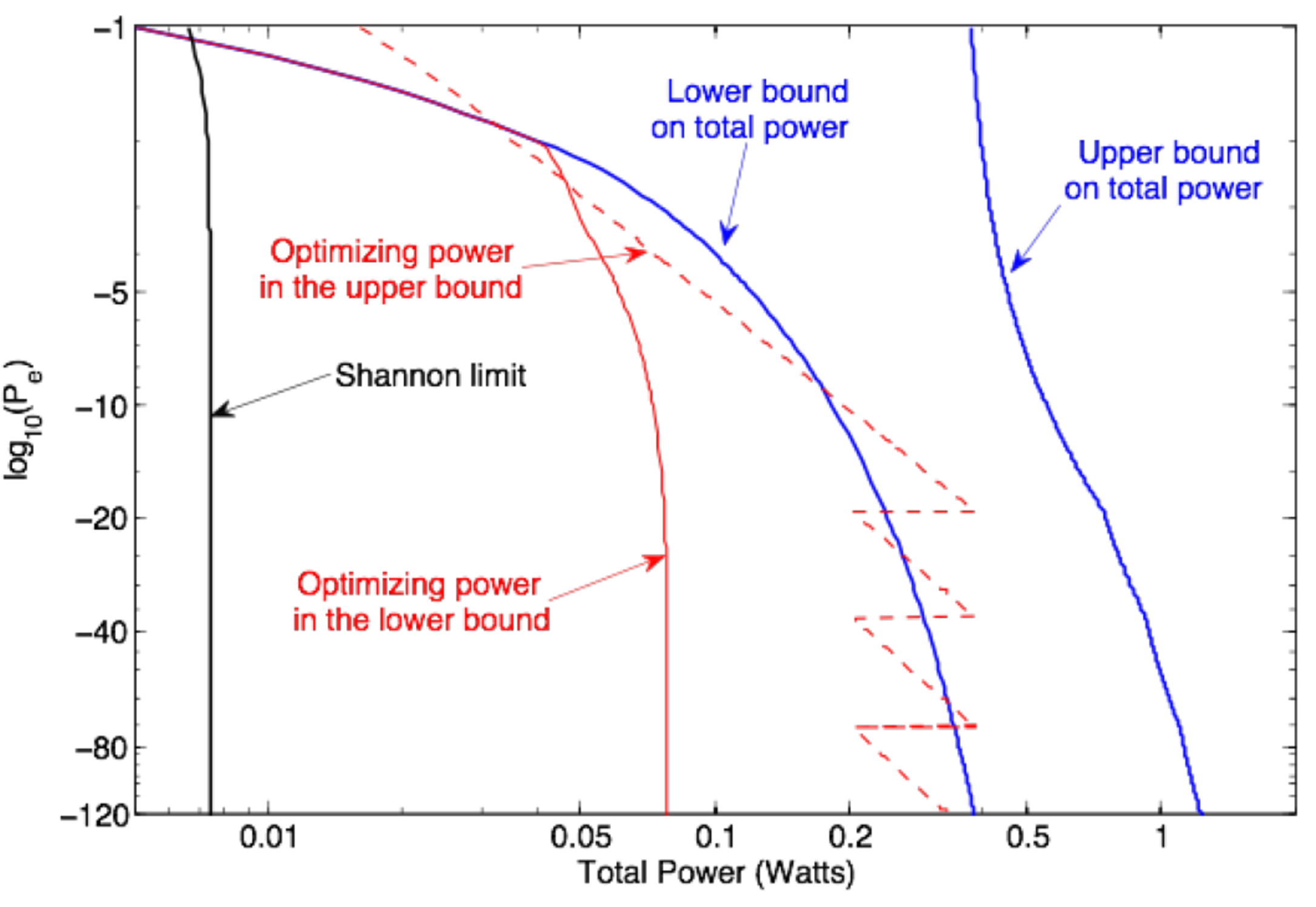}}\\
(a) & (b)
\end{tabular}}
\caption{(a) Thompson's VLSI model of computing. (b) Energy with a model of computation that only considers energy consumed in computational nodes.}
\label{fig:vlsi}
\vspace*{-0.3cm}
\end{figure*}

While insightful, the black-box model has its shortcomings: because it lumps together all of the power used in processing the signal, the black-box model does not yield much insight into code choice or decoder design. Does the code-choice matter? Recent empirical work has shown that energy consumption of the decoding circuitry changes substantially with the choice of the code \cite{Ganesan2011}. Thus a more detailed model of circuit energy is needed in order to guide the choice of the communication strategy as well as the circuit design. One model that has commonly been used to understand tradeoffs in circuit computations is the ``VLSI model of computation,'' developed by Thompson in 1980. The model, illustrated in Fig.~\ref{fig:vlsi}(a), was used by Thompson to derive fundamental limits on the tradeoffs between the required total wiring-length and the number of clock-cycles. The underlying idea is: when the computation to be performed is such that a substantial amount of information needs to be moved around on the circuit, one has to either make wires long so that information can be communicated farther in the same clock-cycle, or increase the number of clock-cycles. These wiring and clock-cycle tradeoffs provide an approximate understanding of the required energy: if we assume that each wire is used in each clock-cycle, the product of the total wiring-length and the number of clock-cycles yields a bound on the required energy. Even this approximate understanding has been made more accurate in the more recent ``information-friction'' model \cite{ISIT13Paper}, which also generalizes the VLSI model so that it is applicable to asynchronous models of computation, and to other substrates of computation (e.g., circuit links made of carbon nanotubes, optical fibers, physical matter transport, or wireless, or even axons that connect neurons in the nervous system).

The VLSI model has been used to understand energy and complexity of encoding and decoding in the system. Focusing on the energy consumed in the computational nodes at the decoder (and ignoring the wiring energy), references \cite{grover, greencodes, JSAC11Paper} showed that the required number of clock-cycles at the decoder (implemented in the VLSI model) diverges to infinity as $P_e\to 0$, and as the communication rate approaches the channel capacity. This is used to obtain a lower bound on total energy. It turns out that to optimize this lower bound, the optimal strategy is to communicate at a constant gap from the Shannon limit as $P_e\to 0$. Numerical evaluation of regular LDPC codes shows that they can achieve order optimal total energy in this model of energy consumption; see Fig.~\ref{fig:vlsi}(b).

However, the empirical work in \cite{Ganesan2011} showed that the \textit{wiring} energy (ignored in \cite{grover, greencodes, JSAC11Paper}) can be a significant fraction of the circuit energy, and can further change substantially with change in the code design (\cite{Ganesan2011} uses LDPC codes where the code-degree and code-length is kept constant but the code-girth is varied). Using the information-friction model to abstract the wiring energy, \cite{ISIT13Paper} (and the ensuing work in \cite{BlakeKschischang}) showed that the wiring energy diverges to infinity much faster than the node energy, an observation consistent with that of circuit practitioners. Further, the total-energy optimal communication strategy is to communicate farther and farther away from the channel capacity as the error-probability converges to zero, and thus keeping the rate close to capacity can actually increase the total power.

Thus the total energy minimizing communication strategy, as well as the scaling limits on total energy as $P_e\to 0$, change drastically when circuit energy is also incorporated. We believe it is important to model and understand the effects of circuit energy (both node and wiring energy) in the context of energy-harvesting as well for the following reasons: Ignoring circuit energy can yield optimistic results and strategies that may significantly underestimate total energy requirements in practice, and be highly inefficient in a total (transmit + circuit) energy sense. In order to understand the energy required in a computation, it is not sufficient to compute the Turing complexity (i.e., count the number of operations) or node energy. Understanding wiring energy (i.e., the energy required to move information between circuit elements) can be key because it tends to dominate the node energy in asymptotics. Thompson's VLSI model, and ensuing improvements, can prove useful in estimating energy requirements of circuitry in an order sense, which can be extremely helpful in deciding the communication strategy. Energy-harvesting circuit might itself need energy, and when systems use joint information and energy transfer, there might be a further increase in cost for circuit energy in order to harvest energy from the same signal. Models for energy-harvesting circuits can thus reveal much about performance of these systems in the real world.

\section{Conclusion and Forward Look}
\label{conclusion}
In this article, we have summarized recent advances that have taken place in the broad area of energy harvesting wireless communication networks. We have covered a variety of topics ranging from information-theoretic physical layer performance limits to scheduling policies and medium access control protocols, as well as the newly emerging paradigm of energy transfer and cooperation that occur in addition or simultaneous with information transfer for such networks. Models and results under a variety of network structures, those with single and multiple hops as well as small and large scale have been addressed. We have also presented models for total energy consumption.

It is worth noting that energy harvesting wireless networks simultaneously present new theoretical challenges and those that stem from physical phenomena and practical concerns. As such, the area provides a rich set of possibilities for obtaining design insights from mathematical formulations which take practical considerations into account. These considerations include such physical properties as storage imperfections, consumption models, processing costs, as well as realistic modeling such as causal energy harvesting profiles. Additionally, the area of energy and information transfer provides exciting possibilities to further adapt the network operation
and improve its performance. The possible improvement therein is closely tied to the efficiency of energy transfer and hence to the device and circuit technologies, connecting the theory again to the real world. To this end, we conclude by stating that the future challenges for energy harvesting wireless networks lie not only in advancements in various layers of network design starting from signal processing and communications physical layer all the way to the networking layer, but also in embracing the truly interdisciplinary nature of the energy harvesting wireless networks integrating with the advances from circuits and devices that harvest and transfer energy.

\bibliographystyle{ieeetr}

\begin{IEEEbiography}{Sennur Ulukus}
(S'90-M'98) is a Professor of Electrical and Computer Engineering at the University of Maryland at College Park, where she also holds a joint appointment with the Institute for Systems Research (ISR). Prior to joining UMD, she was a Senior Technical Staff Member at AT\&T Labs-Research. She received her Ph.D. degree in Electrical and Computer Engineering from Wireless Information Network Laboratory (WINLAB), Rutgers University, and B.S. and M.S. degrees in Electrical and Electronics Engineering from Bilkent University. Her research interests are in wireless communication theory and networking, network information theory for wireless communications, signal processing for wireless communications, information theoretic physical layer security, and energy harvesting communications.

Dr. Ulukus received the 2003 IEEE Marconi Prize Paper Award in Wireless Communications, an 2005 NSF CAREER Award, the 2010-2011 ISR Outstanding Systems Engineering Faculty Award, and the 2012 George Corcoran Education Award. She served as an Associate Editor for the IEEE Transactions on Information Theory (2007-2010) and IEEE Transactions on Communications (2003-2007). She served as a Guest Editor for the IEEE Journal on Selected Areas in Communications for the special issue on wireless communications powered by energy harvesting and wireless energy transfer (2015), Journal of Communications and Networks for the special issue on energy harvesting in wireless networks (2012), IEEE Transactions on Information Theory for the special issue on interference networks (2011), IEEE Journal on Selected Areas in Communications for the special issue on multiuser detection for advanced communication systems and networks (2008). She served as the TPC co-chair of the 2014 IEEE PIMRC, Communication Theory Symposium at 2014 IEEE Globecom, Communication Theory Symposium at 2013 IEEE ICC, Physical-Layer Security Workshop at 2011 IEEE Globecom, Physical-Layer Security Workshop at 2011 IEEE ICC, 2011 Communication Theory Workshop (IEEE CTW), Wireless Communications Symposium at 2010 IEEE ICC, Medium Access Control Track at 2008 IEEE WCNC, and Communication Theory Symposium at 2007 IEEE Globecom. She was the Secretary of the IEEE Communication Theory Technical Committee (CTTC) in 2007-2009.
\end{IEEEbiography}

\begin{IEEEbiography}{Aylin Yener} (S'91-M'00-SM'13-F'14) received the B.Sc. degree in electrical and electronics engineering, and the B.Sc. degree in physics, from Bogazici University, Istanbul, Turkey; and the M.S. and Ph.D. degrees in electrical and computer engineering from Wireless Information Network Laboratory (WINLAB), Rutgers University, New Brunswick, NJ. She is a professor of Electrical Engineering at The Pennsylvania State University, University Park, PA since 2010, where she joined the faculty as an assistant professor in 2002. During the academic year 2008-2009, she was a Visiting Associate Professor with the Department of Electrical Engineering, Stanford University, CA. Her research interests are in information theory, communication theory and network science with recent emphasis on green communications and information security.  She received the NSF CAREER award in 2003, the best paper award in Communication Theory in the IEEE International Conference on Communications in 2010, the Penn State Engineering Alumni Society (PSEAS) Outstanding Research Award in 2010, the IEEE Marconi Prize paper award in 2014, the PSEAS Premier Research Award in 2014, and the Leonard A. Doggett Award for Outstanding Writing in Electrical Engineering at Penn State in 2014.

Dr. Yener has been on the board of governors of the IEEE Information Theory Society as its treasurer (2012-2014). She served as the student committee chair for the IEEE Information Theory Society 2007-2011, and was the co-founder of the Annual School of Information Theory in North America co-organizing the school in 2008, 2009 and 2010. She was a technical (co)-chair for various symposia/tracks at IEEE ICC, PIMRC, VTC, WCNC and Asilomar (2005-2014), and served as an editor for IEEE Transactions on Communications (2009-2012), an editor and an editorial advisory board member for IEEE Transactions on Wireless Communications (2001-2012), a guest editor for IEEE Transactions on Information Forensics and Security (2011) and a guest editor for IEEE Journal on Selected Areas in Communications (2015).
\end{IEEEbiography}

\begin{IEEEbiography}{Elza Erkip}
(S'93-M'96-SM'05-F'11) received the B.S. degree in electrical and electronics engineering from Middle East Technical University, Ankara, Turkey, and the M.S. and Ph.D. degrees in electrical engineering from Stanford University, Stanford, CA, USA. Currently, she is a Professor of electrical and computer engineering with New York University Polytechnic School of Engineering, Brooklyn, NY, USA. Her research interests are in information theory, communication theory, and wireless communications.

Dr. Erkip is a member of the Science Academy Society of Turkey and is among the Thomson Reuters 2014 Edition of Highly Cited Researchers. She received the NSF CAREER award in 2001, the IEEE Communications Society Stephen O. Rice Paper Prize in 2004, the IEEE ICC Communication Theory Symposium Best Paper Award in 2007, and the IEEE Communications Society Award for Advances in Communication in 2013. She co-authored a paper that received the IEEE International Symposium on Information Theory Student Paper Award in 2007. Currently, she is a Member of the Board of Governors of the IEEE Information Theory Society and a Guest Editor of the IEEE JOURNAL ON SELECTED AREAS IN COMMUNICATIONS. Dr. Erkip was a Distinguished Lecturer of the IEEE Information Theory Society from 2013 to 2014, an Associate Editor of the IEEE TRANSACTIONS ON INFORMATION THEORY from 2009 to 2011, an Associate Editor of the IEEE TRANSACTIONS ON COMMUNICATIONS from 2006 to 2008, a Publications Editor of the IEEE TRANSACTIONS ON INFORMATION THEORY from 2006 to 2008 and a Guest Editor of the IEEE SIGNAL PROCESSING MAGAZINE in 2007. She was a General Chair for the IEEE International Symposium of Information Theory in 2013, a Technical Program Chair for the International Symposium on Modeling and Optimization in Mobile, Ad Hoc, and Wireless Networks (WiOpt) in 2011, a Technical Program Chair for the IEEE GLOBECOM Communication Theory Symposium in 2009, the Publications Chair for the IEEE Information Theory Workshop, Taormina, in 2009, the Technical Area Chair for the MIMO Communications and Signal Processing track of Asilomar Conference on Signals, Systems, and Computers in 2007, and a Technical Program Chair for the IEEE Communication Theory Workshop in 2006.
\end{IEEEbiography}

\begin{IEEEbiography}
{Osvaldo Simeone} received the M.Sc. degree (with honors) and the Ph.D. degree in information engineering from Politecnico di Milano, Milan, Italy, in 2001 and 2005, respectively. He is currently with the Center for Wireless Communications and Signal Processing Research (CWCSPR), New Jersey Institute of Technology (NJIT), Newark, where he is an Associate Professor. His current research interests concern wireless communications, information theory and machine learning. Dr. Simeone is a co-recipient of Best Paper Awards of the IEEE SPAWC 2007
and IEEE WRECOM 2007. He currently serves as an Editor for IEEE Transactions on Information Theory.
\end{IEEEbiography}

\begin{IEEEbiography}
{Michele Zorzi} (S'89-M'95-SM'98-F'07) received his Laurea and PhD degrees in electrical engineering from the University of Padova in 1990 and 1994, respectively. During academic year 1992-1993 he was on leave at UCSD, working on multiple access in mobile radio networks. In 1993 he joined the faculty of the Dipartimento di Elettronica e Informazione, Politecnico di Milano, Italy. After spending three years with the Center for Wireless Communications at UCSD, in 1998 he joined the School of Engineering of the University of Ferrara, Italy, where he became a professor in 2000. Since November 2003 he has been on the faculty of the Information Engineering Department at the University of Padova. His present research interests include performance evaluation in mobile communications systems, random access in wireless networks, ad hoc and sensor networks, Internet-of-Things, energy constrained communications protocols, cognitive networks, and underwater communications and networking.

He was Editor-In-Chief of IEEE Wireless Communications from 2003 to 2005 and Editor-In-Chief of the IEEE Transactions on Communications from 2008 to 2011, and is the founding Editor-In-Chief of the IEEE Transactions on Cognitive Communications and Networking. He has also been an Editor for several journals and a member of the Organizing or the Technical Program Committee for many international conferences, as well as guest editor for special issues in IEEE Personal Communications (``Energy Management in Personal Communications Systems''), IEEE Network (``Video over Mobile Networks'') and IEEE Journal on Selected Areas in Communications (``Multimedia Network Radios'' and ``Underwater Wireless Communications and Networking''). He served as a Member-at-Large of the Board of Governors of the IEEE Communications Society from 2009 to 2011, and is currently its Director of Education.
\end{IEEEbiography}

\begin{IEEEbiography}
{Pulkit Grover} (Ph.D. UC Berkeley'10, B.Tech.'03, M.Tech.'05 IIT Kanpur) is an assistant professor at CMU. Prior to joining CMU in 2013, he was a postdoctoral researcher at Stanford. He is interested in interdisciplinary research directed towards developing a science of information for making decentralized sensing, communication and computing systems (including biomedical systems) energy-efficient and stable. He is the recipient of the 2010 best student paper award at the IEEE Conference in Decision and Control (CDC); a finalist at the 2010 best student paper award at the IEEE International Symposium on Information Theory (ISIT); the 2011 Eli Jury Award from UC Berkeley; the 2012 Leonard G. Abraham best paper award from the IEEE Communications Society; a 2014 best paper award at the International Symposium on Integrated Circuits (ISIC); and a 2014 NSF CAREER award.
\end{IEEEbiography}

\begin{IEEEbiography}
{Kaibin Huang} (M'08-SM'13) received the B.Eng. (first-class hons.) and the M.Eng. from the National University of Singapore in 1998 and 2000, respectively, and the Ph.D. degree from The University of Texas at Austin (UT Austin) in 2008, all in electrical engineering.

Since Jan. 2014, he has been an assistant professor in the Dept. of Electrical and Electronic Engineering (EEE) at The University of Hong Kong. He is an adjunct professor in the School of EEE at Yonsei University in S. Korea. He used to be a faculty member in the Dept. of Applied Mathematics (AMA) at the Hong Kong Polytechnic University (PolyU) and the Dept. of EEE at Yonsei University. He had been a Postdoctoral Research Fellow in the Department of Electrical and Computer Engineering at the Hong Kong University of Science and Technology from Jun. 2008 to Feb. 2009 and an Associate Scientist at the Institute for Infocomm Research in Singapore from Nov. 1999 to Jul. 2004. His research interests focus on the analysis and design of wireless networks using stochastic geometry and multi-antenna techniques.

He frequently serves on the technical program committees of major IEEE conferences in wireless communications. He chairs the Comm. Theory Symp. of IEEE GLOBECOM 2014 and the Adv. Topics in Wireless Comm. Symp. of IEEE/CIC ICCC 2014 and has been the technical co-chair for IEEE CTW 2013, track chairs for IEEE PIMRC 2015, IEE VTC Spring 2013, Asilomar 2011 and IEEE WCNC 2011. He is a guest editor for the IEEE Journal on Selected Areas in Communications, an editor for the IEEE Transactions on Wireless Communications, IEEE Wireless Communications Letters and also IEEE/KICS Journal of Communication and Networks. He is an elected member of the SPCOM Technical Committee of the IEEE Signal Processing Society. Dr. Huang received the Outstanding Teaching Award from Yonsei, Motorola Partnerships in Research Grant, the University Continuing Fellowship from UT Austin, and a Best Paper Award from IEEE GLOBECOM 2006 and PolyU AMA in 2013.
\end{IEEEbiography}

\end{document}